\documentclass[fleqn,useAMS,usenatbib]{mnras}

\usepackage{amsmath}
\usepackage{amssymb}
\usepackage{graphicx}
\usepackage{times}
\usepackage{hyperref}

\usepackage[flushleft]{threeparttable}

\usepackage{xspace}
\newcommand{\ketju}{\textsc{ketju}\xspace}

\title[BH binary model]
  {Modelling the accretion and feedback of supermassive black hole binaries in gas-rich galaxy mergers}

\author[S. Liao et al.]
{Shihong Liao$^{1}$\thanks{Email: shihong.liao@helsinki.fi}, Peter H. Johansson$^{1}$, Matias Mannerkoski$^{1}$, 
Dimitrios Irodotou$^{1}$, \newauthor Francesco Paolo Rizzuto$^{1}$, Stuart McAlpine$^{2}$, Antti Rantala$^{3}$, Alexander Rawlings$^{1}$, \newauthor Till Sawala$^{1,4}$
\\
$^1$Department of Physics, University of Helsinki, Gustaf Hällströmin katu 2, FI-00014 Helsinki, Finland\\
$^2$The Oskar Klein Centre, Department of Physics, Stockholm University, Albanova University Center, 106 91 Stockholm, Sweden \\
$^3$Max-Planck-Institut f\"ur Astrophysik, Karl-Schwarzchild-Str 1, D-85748 Garching, Germany\\
$^4$Institute for Computational Cosmology, Durham University, South Road, Durham DH1 3LE, United Kingdom\\
}

\begin{document}



\maketitle

\label{firstpage}

\begin{abstract}
We introduce a new model for the accretion and feedback of supermassive black hole (SMBH) binaries to the \ketju code, which enables us to resolve the evolution of SMBH binaries down to separations of tens of Schwarzschild radii in gas-rich galaxy mergers. Our subgrid binary accretion model extends the widely used Bondi--Hoyle--Lyttleton accretion into the binary phase and incorporates preferential mass accretion onto the secondary SMBH, which is motivated by results from small-scale hydrodynamical circumbinary disc simulations. We perform idealised gas-rich disc galaxy merger simulations using pure thermal or pure kinetic active galactic nuclei (AGN) feedback. Our binary accretion model provides more physically motivated SMBH mass ratios, which are one of the key parameters for computing gravitational wave (GW) induced recoil velocities. The merger time-scales of our simulated SMBH binaries are in the range $t_{\rm merge}{\sim} 10$--$400$ Myr. Prograde in-plane equal-mass galaxy mergers lead to the shortest merger time-scales, as they experience the strongest starbursts, with the ensuing high stellar density resulting in a rapid SMBH coalescence. Compared to the thermal AGN feedback, the kinetic AGN feedback predicts longer merger time-scales and results in more core-like stellar profiles, as it is more effective in removing gas from the galaxy centre and quenching star formation. This suggests that the AGN feedback implementation plays a critical role in modelling SMBH coalescences. Our model will be useful for improving the modelling of SMBH mergers in gas-rich galaxies, the prime targets for the upcoming LISA GW observatory.
\end{abstract}


\section{Introduction}\label{sec:intro}

Observations have revealed that supermassive black holes (SMBHs, or BHs for short), with masses ranging from ${\sim} 10^{5}$ to ${\sim} 10^{10}~{\rm M}_{\sun}$, are ubiquitous at the centres of massive galaxies and their masses tightly correlate with the properties of their host galaxies, suggesting a co-evolution picture of SMBHs and their host galaxies \citep[see e.g.][for a review]{Kormendy2013}. The interplay between SMBHs and their host galaxies motivated by this co-evolution picture has become one of the key ingredients in modern galaxy formation models \citep[see][for a review]{Naab2017}. SMBHs grow by accreting gas from their host galaxies; at the same time they release energy to the surrounding gas, which is named active galactic nuclei (AGN) feedback, leading to self-regulated SMBH growth and affecting the overall properties of host galaxies. AGN feedback has been shown to be a crucial mechanism in quenching massive galaxies and reproducing the bright end of the observed galaxy luminosity function \citep[e.g.][]{Bower2006,Croton2006}.

In the current standard cosmological framework based on the cold dark matter paradigm, structures (galaxies) grow hierarchically, i.e. larger structures form through the continuous merging of smaller ones \citep*[see e.g.][for reviews]{Mo2010,Frenk2012}. During a merger between two massive galaxies, the evolution of the two central SMBHs can be decomposed into three major stages \citep*{Begelman1980}: (i) at kpc scale separation, due to dynamical friction, the SMBHs lose their orbital energy and angular momentum, sink towards the centre of the galaxy merger remnant, and form a gravitationally bound binary; (ii) at pc scales, the SMBH binary continues to shrink by interacting with gas (e.g. circumbinary disc, \citealt{Haiman2009}) and individual stars (e.g. slingshot interactions between the binary and stars, \citealt{Hills1980}); (iii) at mpc scales, the SMBH binary dynamics is dominated by the gravitational wave (GW) emission until coalescence.

The GWs generated from SMBH coalescence are a major target for current and future low-frequency GW detectors such as the Pulsar Timing Array \citep[PTA, frequency range of ${\sim} 10^{-9}$ to ${\sim} 10^{-6}$ Hz, see e.g.][for a review]{Hobbs2017} and space-based GW observatories like the Laser Interferometer Space Antenna \citep[LISA, frequency range of $10^{-4}$ to $10^{-1}$ Hz,][]{Amaro-Seoane2017} and TianQin \citep[$10^{-3}$ to $10^{-1}$ Hz,][]{Luo2016}. In order to offer better theoretical predictions for these GW experiments and to maximise their scientific returns, it is of paramount importance to improve the modelling of BH formation and evolution in a cosmological context of galaxy formation, while simultaneously resolving the small-scale dynamics of BH binaries \citep[e.g.][]{Amaro-Seoane2022}.

In cosmological galaxy formation simulations, because of the limited spatial and mass resolution, the sub-pc-scale BH accretion disc is not resolved and BH evolution is modelled in a subgrid approach as follows:

(i) {\it BH accretion and AGN feedback.} BH accretion is typically implemented using the Bondi--Hoyle--Lyttleton \citep[BHL,][]{Hoyle1939,Bondi1944,Bondi1952} accretion model. The BHL model was first introduced into galaxy formation simulations by \citet{DiMatteo2005} and \citet{Springel2005BH} and it has been adopted in many later large simulation projects such as  Illustris \citep{Vogelsberger2013}, EAGLE \citep{Schaye2015}, Illustris TNG \citep{Weinberger2017,Pillepich2018}, Auriga \citep{Grand2017} amongst others. In addition, there are also other accretion models, e.g. the gravitational torque-driven accretion model \citep{Hopkins2011,Angles-Alcazar2017,Dave2019}, the transport of angular momentum model \citep[][]{Debuhr2011,Debuhr2012}, and the accretion disc particle model \citep{Power2011}. In different implementations of AGN feedback, BHs can heat the surrounding gas via thermal feedback \citep[e.g.][]{Springel2005BH,Booth2009,Johansson2009ApJ,Teyssier2011,Schaye2015,Blank2019,Tremmel2019,Ni2022}, or kick surrounding gas elements by jets/winds through kinetic feedback, or adopt multiple feedback mechanisms \citep[e.g.][]{Choi2012,Debuhr2012,Dubois2012,Barai2014,Costa2014,Angles-Alcazar2017,Weinberger2017,Dave2019}. See \citet{Habouzit2021} and \citet{Habouzit2022} for comparisons of the AGN feedback models in some large-scale cosmological simulations.

(ii) {\it BH dynamics.} Due to the softened gravitational interaction, small-scale BH dynamics is not resolved in standard galaxy formation simulations, especially for low-mass BHs and at the scales of the BH binary phase. Instead, BHs are typically repositioned to the local potential minima assuming that the unresolved dynamical friction can efficiently keep BHs near galaxy centres \citep[e.g.][]{Springel2005BH,Johansson2009ApJ,Bahe2022}. In addition, the two BHs are merged instantaneously if their separation and relative velocity satisfy some given criteria. For example, in \citet{Springel2005BH}, two BHs are merged if their separation is less than the BH smoothing length (usually kpc or sub-kpc scales) which is defined using the smoothed particle hydrodynamics approach and their relative velocity is below the local sound speed.

As a result, the three-stage BH coalescence process is not fully resolved in standard cosmological simulations. Although there have been some improvements in resolving BH dynamics at kpc or sub-kpc scales by introducing dynamical friction subgrid modelling to replace the repositioning scheme \citep[e.g.][]{Tremmel2015,Pfister2019,Chen2022,Ma2022}, the evolution beyond the formation of a bound BH binary and the resulting GW predictions usually have to be modelled by post-processing the simulation data with semi-analytical prescriptions \citep[e.g.][]{Salcido2016,Kelley2017,Katz2020,Volonteri2020,Chen2022Astrid,Li2022}.

Recently, there have also been attempts to resolve the entire BH coalescence process in cosmological simulations. \citet{Khan2016} selected two merging disc galaxies in a group environment from the Argo cosmological zoom-in hydrodynamical simulation \citep{Feldmann2015} and resimulated the evolution of the very central region with their direct $N$-body code \textsc{$\phi$-gpu} until the two BHs coalesce. To do so, they had to manually convert all gas particles into stars before the onset of the direct $N$-body simulation \citep[see similar approaches in][]{Khan2012,Khan2018}. Moreover, they did not include BH accretion and AGN feedback modelling in both the parent hydrodynamical simulation and the resimulations. These over-simplified approximations make their approach not fully self-consistent, and how these simplifications affect the simulated BH dynamics is unknown. Recently, \citet{Mannerkoski2021} and \citet{Mannerkoski2022} performed cosmological zoom-in simulations of galaxy groups which resolve BH dynamics down to separations of tens of Schwarzschild radii while simultaneously modelling the galaxy formation processes (e.g. cooling, star formation, stellar and AGN feedback). These more self-consistent simulations enable us to study the evolution of both BH binary and triplet systems, the impacts of GW recoiling velocity on BH-galaxy scaling relations, and the GW signatures from BH mergers.

In the simulations of \citet{Mannerkoski2021} and \citet{Mannerkoski2022}, given that most of the BH binaries are in a gas-poor environment and their accretion rates are in general low, the authors still adopted the traditional BHL accretion and isotropic thermal feedback model, which was originally introduced for single BHs. When simulating BH binaries in gas-rich galaxy mergers, the gas accretion rates can be significantly higher, and the BH model should be improved to properly take into account the BH binary phase. Note that this is particularly important for modelling LISA targets, i.e. BHs with masses in the range of $10^{3} - 10^{7}~{\rm M}_{\sun}$ up to $z \sim 20$ \citep{Amaro-Seoane2022}, as these low-mass and high-redshift BH mergers usually reside in gas-rich galaxies. 

During the evolution of a BH binary, the gas from the circumnuclear disc \citep[e.g.][]{Escala2005,Dotti2007,Mayer2007} forms a circumbinary disc centred at the binary centre-of-mass, leading to a different accretion behaviour compared to the single BH case \citep[see][for a review]{Lai2022}. For example, small-scale viscous hydrodynamical simulations of circumbinary discs have revealed that during the BH binary phase, there is a preferential accretion onto the secondary BH, i.e. the long-term accretion rate of the secondary BH is higher than that of the primary BH \citep[e.g.][]{Artymowicz1996,Hayasaki2007,Farris2014,Duffell2020,Munoz2020}. In contrast, in the BHL model, the BH accretion rate is proportional to the squared BH mass and thus it is higher for the more massive BH. These two models lead to two opposite outcomes: a binary in the circumbinary disc accretion model tends to be equal-mass while a binary in the BHL model evolves towards a larger mass ratio. This has an important impact on the simulation predictions since the BH mass ratio is one of the key parameters that affects the GW-induced recoil velocity \citep[][]{Campanelli2007,Baker2008,Lousto2009,Zlochower2015}. Therefore an improved BH accretion and feedback model needs to be developed for simulating BH binaries in a gas-rich environment.

In this work, we introduce a BH binary accretion and feedback model which incorporates the results from small-scale circumbinary disc simulations in a subgrid approach, and use gas-rich disc galaxy mergers to test and demonstrate the behaviour of the model. 
The structure of our paper is as follows. In Section~\ref{sec:num_models}, we describe the details of our numerical code and the galaxy formation subgrid model (i.e. gas cooling, star formation, and stellar feedback). In Section~\ref{sec:bh_model}, we introduce the BH accretion and feedback models. The simulation initial conditions and the simulation details are described in Section~\ref{sec:sim}. Our results are presented in Section~\ref{sec:res}. We discuss the caveats and future improvements of our model in Section~\ref{sec:dis} and finally conclude in Section~\ref{sec:con}.

\section{Numerical code}\label{sec:num_models}

\subsection{The \ketju code}\label{subsec:ketju_code}

The \ketju code \citep{Rantala2017} extends the widely used \textsc{gadget-3} code \citep{Springel2005GADGET2} by replacing the standard leapfrog integration with the high-accuracy algorithmically regularised \textsc{Mstar} integrator \citep{Rantala2020} in selected regions around each BH. All gravitational interactions in the regularised regions between BHs, and BHs and star particles, are computed without gravitational softening, whereas star-star interactions are softened in order to avoid energy errors when particles enter and exit the regularised regions. The standard gadget leapfrog integrator is also used for integrating the centre-of-mass motion of the regularised regions, as well as all the other simulation particles outside the regularised regions. In order to model the effects of general relativity on the motion of the BHs, the \ketju code also includes  post-Newtonian (PN) corrections for BH-BH interactions up to the order of PN3.5 \citep{Mora2004}.

\subsection{Hydrodynamics}\label{sec:hydro}

The hydrodynamics is modelled using the modern \textsc{sphgal} smooth particle hydrodynamics (SPH) implementation developed by \citet{Hu2014}. In this model, the SPH calculation is performed using the pressure-entropy formulation utilising a Wendland $C^4$ kernel with $N_{\rm ngb} = 100$ neighbouring particles, and also includes a new artificial viscosity scheme with a stronger viscosity coefficient limiter \citep{Cullen2010} and the artificial conduction of thermal energy \citep{Read2012}. Together, these improvements help to better resolve the fluid mixing at contact discontinuities and prevent the development of feedback-induced viscous instabilities in isolated disc galaxy simulations. 

In addition, the \textsc{sphgal} model also uses sophisticated time-step control by restricting the time-steps of neighbouring SPH particles within a factor of few (we use $4$ in this study) for particles undergoing strong shocks \citep{Saitoh2009}. In addition, the \textsc{sphgal} model adopts the time-step limiter proposed by \citet{Durier2012} when modelling feedback processes. Thus, when BHs or stars inject thermal or kinetic feedback energy to SPH particles, the inactive energy-receiving SPH particles will immediately become active and shorten their time-steps. This is necessary to ensure that the system reacts to energy inputs promptly and achieves proper energy conservation. 

\subsection{Gas cooling, star formation, and stellar feedback}\label{subsection:stellar_feedback}

Our adopted model for gas cooling, star formation and stellar feedback was originally developed by \citet{Scannapieco2005,Scannapieco2006} and later improved by \citet{Aumer2013} and \citet{Nunez2017}. It has been tested and used both in isolated galaxy and merger simulations \citep{Eisenreich2017,Lahen2018}, and more recently also in cosmological zoom-in simulations  \citep{Mannerkoski2021,Mannerkoski2022}.

In this model, the abundances of eleven chemical elements (H, He, C, N, O, Ne, Mg, Si, S, Ca, and Fe) are traced for every gas and star particle in the simulation, with the cooling rate for each gas particle depending on its current temperature, density, and chemical composition. The cooling rate tables are adopted from \citet*{Wiersma2009} assuming that the gas is optically thin and in ionisation equilibrium embedded in the redshift-dependent UV/X-ray background from quasars and galaxies \citep{Haardt2001} and the cosmic microwave background. In this study we use the cooling table calculated with the $z=0$ background with a lower temperature limit of $T=10^{4} \ \rm K$. 

During the simulation, the metal abundances for gas particles evolve due to the metal enrichment from stellar feedback and the turbulent diffusion of metals \citep{Aumer2013}. Note that as gas particles keep receiving metals from stars during the simulation, the masses of some gas particles can reach twice the initial mass. To keep a relatively uniform mass resolution, we split such an over-massive gas particle into two gas particles and record the splitting history so that we can trace the evolution of baryonic particles in our simulations.

Stars form stochastically from gas particles with densities $\rho_{\rm gas} \geq 2.2 \times 10^{-24}~{\rm g}~{\rm cm}^{-3}$ (or equivalently hydrogen number densities $n_{\rm H} \geq 1~{\rm cm}^{-3}$) and with temperatures $T \leq 12000~{\rm K}$ which reside in convergent flows, i.e. the velocity divergence $\nabla \cdot  \mathbfit{v}_{\rm gas} \leq 0$. The probability of converting a gas particle into a star particle is given by
\begin{equation}
    p_{\rm SF} = \epsilon_{\rm SFR} \frac{\Delta t}{t_{\rm dyn}} = \epsilon_{\rm SFR} \Delta t \sqrt{4 \upi G \rho_{\rm gas}},
\end{equation}
where $\epsilon_{\rm SFR}=0.02$ is the star formation efficiency, $\Delta t$ is the size of the timestep, $t_{\rm dyn}$ is the dynamical time, and $G$ is the gravitational constant.

We consider the stellar feedback from the explosions of type Ia (SNIa) and type II (SNII) supernovae and the slow winds from asymptotic giant branch (AGB) stars. When a star particle is flagged for feedback, it enriches, heats (thermal feedback), and kicks (kinetic feedback) the closest 10 neighbouring gas particles. The fractions of the metals and energy that a gas neighbour receives are weighted by the SPH smoothing kernel of the star particle. To avoid adding feedback mass and energy to gas particles at unphysically large distances when there are not enough nearby gas particles, we introduce a maximum radius of $r_{{\rm max}, \star} = 2~{\rm kpc}$ to the neighbour search \citep[see][for a similar implementation of a maximum search radius]{Hopkins2018}.

Each star particle represents a stellar population with a \citet{Kroupa2001} initial mass function. To estimate the SNII rates, we assume that stars more massive than $8~{\rm M}_{\sun}$ end their lives as SNII. As massive stars have fairly short lifetimes ($\la 10^{7}~{\rm yr}$), we simply assume that after the formation of a star particle, all SNII in this star particle explode at the same age of $\tau_{\rm SNII} = 3~{\rm Myr}$. In contrast, CO white dwarf binary systems, which are the SNIa progenitors, can explode at very different ages. For a star particle in the simulation, the first SNIa explodes at the age of $\tau_{\rm SNIa,~min} = 50~{\rm Myr}$, and the following SNIa feedback events are performed every $50~{\rm Myr}$ until a maximum time period of $\tau_{\rm SNIa,~max} = 10 ~ {\rm Gyr}$ is reached. To estimate the SNIa rates, we adopt a delay time distribution, which declines with the age of the star particle as $\tau^{-1}$, and the normalisation of 2 SNIa per $1000~{\rm M}_{\sun}$ of formed stars following \citet{Maoz2012}. 

When a star particle undergoes a supernova feedback event (SNII or SNIa), mass is ejected with an outflow velocity of $v_{\rm SN} = 4000~{\rm km}~{\rm s}^{-1}$ to the gas neighbours, resulting in a total energy of the supernova feedback event of
\begin{equation}
    E_{\rm SN} = \frac{1}{2} m_{\rm ej} v_{\rm SN}^2,
\label{Eq:SN_ej}
\end{equation}
where $m_{\rm ej}$ is the ejected mass. For SNII and SNIa, the ejected masses (or chemical enrichment) are computed by adopting the yields from \citet{Woosley1995} and \citet{Iwamoto1999} respectively. 

Following the supernova feedback model of \citet{Nunez2017}, the ejecta received by a gas particle is in one of the three supernova remnant phases depending on the distance between the star particle and the receiving gas particle: the momentum-conserving free expansion (FE), the energy-conserving Sedov--Taylor (ST), or the momentum-conserving snowplow (SP) phase. In the FE phase, the ejected mass is much greater than the mass swept out by the supernova remnant blast wave, and the ejecta expands at an almost constant velocity and conserves momentum when impacting the surrounding medium. The supernova feedback is added to the gas particle assuming a perfectly inelastic collision, resulting in a fraction of the energy being transferred in  kinetic form and the remaining energy in a thermal form. In the ST phase, the blast wave expands adiabatically, experiencing only negligible radiative losses, and it can be described by a self-similar solution in which 30 per cent of the ejected energy is kinetic energy, whereas the remaining 70 per cent is thermal energy. In the SP phase, the blast wave experiences significant energy loss due to radiative cooling, and thus the ejected energy decreases as a function of the radial distance. We refer the reader to \citet{Nunez2017} for a detailed description and implementation of this supernova feedback model.

The AGB stellar feedback is implemented similarly to the aforementioned supernova feedback in the FE phase, assuming that the momentum is conserved during the feedback process, and with the same time delay function as the SNIa feedback.
The chemical enrichment for the AGB feedback is computed using the metal-dependent yields from \citet{Karakas2010} and the AGB wind velocity is also modelled using Eq. (\ref{Eq:SN_ej}) but assuming a much lower velocity of $v_{\rm AGB} = 25~{\rm km}~{\rm s}^{-1}$.  

\section{BH model}\label{sec:bh_model}

\subsection{Traditional single BH accretion (gadget)}\label{subsec:single_bh}

In the BHL accretion model \citep{Hoyle1939,Bondi1944,Bondi1952,DiMatteo2005,Springel2005BH}, the mass accretion rate is computed as
\begin{equation}
    \dot{M}_{\rm BHL} = \alpha \frac{4\upi G^2 M_{\rm BH}^2 \rho}{\left(c_{\rm s}^2 + v_{\rm rel}^2\right)^{3/2}},
\end{equation}
where $M_{\rm BH}$ is the BH internal mass, and $\rho$, $c_{\rm s}$, and $v_{\rm rel}$ are the gas density, sound speed, and the magnitude of the relative velocity between the BH and gas at the location of the BH, respectively. We use the same SPH approach as for gas particles to define the BH smoothing length ($h_{\rm BH}$) and estimate the gas properties at the BH location using all SPH particles within $h_{\rm BH}$. The dimensionless boost factor, $\alpha$, is introduced due to the limit of the numerical resolution in galaxy formation simulations. We adopt $\alpha = 25$ in this study.

The BH mass accretion rate is capped by the Eddington accretion rate, i.e.
\begin{equation}\label{eq:bh_accr_rate}
    \dot{M}_{\rm BH} = {\rm min}(\dot{M}_{\rm BHL}, \dot{M}_{\rm Edd}),
\end{equation}
where the Eddington accretion rate is
\begin{equation}
    \dot{M}_{\rm Edd} = \frac{4 \upi G M_{\rm BH} m_{\rm p}}{\epsilon_{\rm r} \sigma_{\rm T} c},
\end{equation}
with $m_{\rm p}$, $\epsilon_{\rm r}$, $\sigma_{\rm T}$, and $c$ being the proton mass, the radiative efficiency, the Thomson cross section, and the speed of light, respectively. In this study, we adopt $\epsilon_{\rm r} = 0.1$.

In the actual implementation, the internal mass of a BH particle is updated at every time step as
\begin{equation}
    M_{\rm BH}(t + \Delta t) = M_{\rm BH}(t) + (1 - \epsilon_{\rm r}) \dot{M}_{\rm BH} \Delta t.
\end{equation}
The factor of $(1 - \epsilon_{\rm r})$ is introduced here as an amount of energy $\Delta E_{\rm r} = L_{\rm r} \Delta t$ has been radiated. Here, the radiated luminosity is $ L_{\rm r} \equiv \epsilon_{\rm r} \dot{M}_{\rm BH} c^2$. The real dynamical mass of a BH particle, which is used in the gravity calculations, tracks its internal mass closely in a probabilistic approach, i.e. at each time step, the probability of a gas particle being swallowed by the BH is
\begin{equation}
    p_i = \frac{w_i \dot{M}_{\rm BH} \Delta t}{\rho},
\end{equation}
where $w_i$ is the SPH kernel weight. For each gas particle, a uniform random number $x_i \in [0, 1]$ is generated, and if $x_i < p_i$, a fraction $(1 - \epsilon_{\rm r})$ of the mass and momentum of this gas particle is added to the BH and the gas particle becomes a ghost particle thereafter.

As the dynamical friction from stars, dark matter, and gas is not well resolved in traditional galaxy formation simulations, it is usually assumed that this dynamical friction is able to efficiently keep the BH near the galaxy centre. Specifically, at every time step, we compute the gravitational potential and the particle-BH relative velocity $v_{\rm pBH}$ for all types of particles within $h_{\rm BH}$, then reposition the BH to the position of the particle with the minimum potential and with $v_{\rm pBH} \leq 0.25 c_{\rm s}$.

When the distance between two BHs is smaller than any of their smoothing lengths and their relative velocity satisfies $v_{\rm BH,rel} \leq 0.5 c_{\rm s}$, these two BHs are merged instantaneously. Therefore, in the traditional galaxy formation simulations, the BH binary phase is not resolved. In this paper we denote the traditional BH model outlined in this subsection as `{\bf gadget}'.

\subsection{Ketju + single BH accretion}\label{subsec:ketju_single_bh}

With the ketju integrator, we are able to resolve the BH dynamics in the binary phase, which further enables us to resolve the complex evolution of many BH mergers and their resulting GW signals in zoom-in galaxy formation simulations \citep{Mannerkoski2021,Mannerkoski2022}. As in our previous zoom-in simulations of galaxy groups, the galaxies were relatively gas-poor and the BH accretion was less important, we adopted the single BH accretion model described in Section \ref{subsec:single_bh} as an approximation, but with the following modifications: (i) The repositioning prescription is not used for BHs in ketju regularised regions as the non-softened gravity between star particles and BHs are calculated within the regularised regions, and the dynamical friction from stars is thus properly modelled. (ii) Two BHs are merged at a distance of 6 times the combined Schwarzschild radius, i.e. $12 G (M_{\rm BH, 1} + M_{\rm BH, 2}) / c^2$ (where $M_{\rm BH,1}$ and $M_{\rm BH,2}$ are the masses of the primary and secondary black holes respectively and $M_{\rm BH, 1} \geq M_{\rm BH, 2}$ is assumed in this paper) as the ketju simulations include PN corrections up to order 3.5 for BH-BH interactions and can resolve the evolution of a BH binary down to the separation regime where the hardening is dominated by GW emission. In this paper, this approach is denoted as `{\bf ketju + single accretion}'.

\subsection{Ketju + binary BH accretion}\label{subsec:binary_bh}

For gas-rich galaxy mergers, BH accretion becomes more important, and the gas accretion during the binary phase has to be carefully modelled. In this subsection, we introduce a BH binary accretion model for the \ketju galaxy formation simulation code.

We adopt a simple switch between the single and binary phase in our model: when a BH is isolated (i.e. it does not form a bound binary with any other BH), we use the `ketju + single accretion' model; when a BH forms a bound binary system with another BH, we use the binary accretion subgrid model which incorporates as an input the results from circumbinary disc simulations (e.g. \citealt{Artymowicz1996,Hayasaki2007,Farris2014,Duffell2020,Munoz2020}; see \citealt{Lai2022} for a recent review). Two nearby BHs are defined as a bound binary when their total orbital energy satisfies
\begin{equation}
    E_{12} = \frac{1}{2}\mu v_{12}^2 - G\frac{M_{\rm BH, 1} M_{\rm BH, 2}}{r_{12}} < 0,
\end{equation}
where $\mu = M_{\rm BH, 1} M_{\rm BH, 2} / (M_{\rm BH, 1} + M_{\rm BH, 2})$ is the reduced mass, and $v_{12}$ and $r_{12}$ are the relative velocity and separation between two BHs, respectively.

\begin{figure} 
\centering\includegraphics[width=200pt]{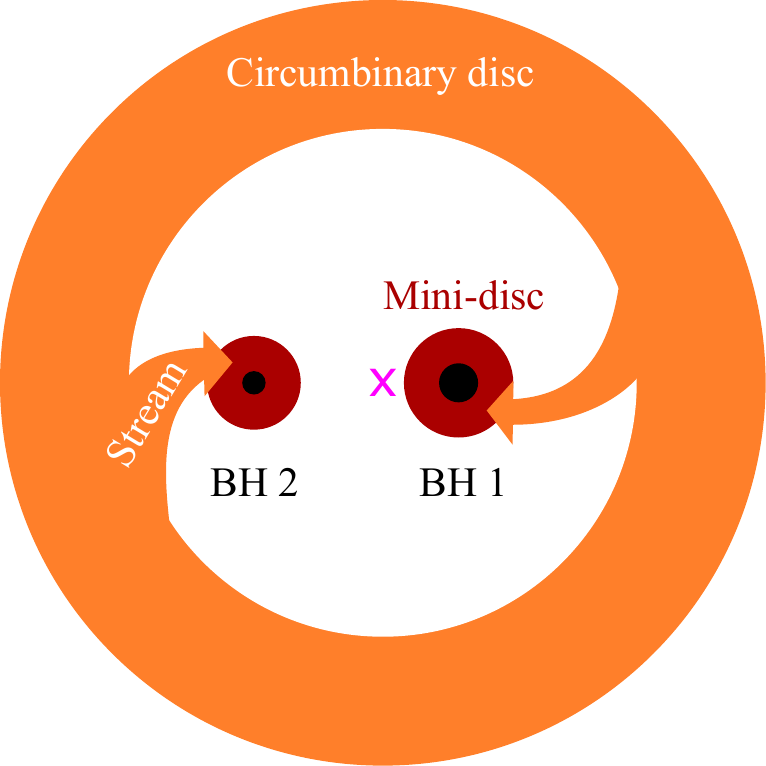}
\caption{Schematic diagram of a circumbinary disc (face-on). The primary and secondary BHs are marked as `BH 1' and `BH 2' respectively. The magenta cross marks the centre-of-mass position of the BH binary.}
\label{fig:circumbinary_disc}
\end{figure}

When two BHs form a bound binary system, the gas from the circumnuclear disc forms a circumbinary disc centred on the binary centre-of-mass (CoM) position (see Fig.~\ref{fig:circumbinary_disc} for a schematic diagram). According to circumbinary disc simulations, due to the tidal torques from the binary, the circumbinary disc develops a central cavity with a radius of roughly twice the binary semi-major axis where the gravitational torque from the binary balances the viscous torque from the disc. Gas streams from the circumbinary disc transport gas into the BH neighbourhood and mini-discs centred on each BH form. As the secondary BH is closer to the inner edge of the circumbinary disc, it can accrete gas from the circumbinary disc more readily and thus it preferentially has a higher accretion rate compared to the primary BH. The consequence of this preferential accretion is that two BHs in a bound binary evolve towards equal-mass.

As the scales of the circumbinary disc are not resolved in cosmological galaxy formation simulations, we incorporate this preferential accretion of the secondary BH into our BH model in a subgrid approach. We first use the BHL accretion formalism to estimate the total accretion rate for the whole binary system (or equivalently the total gas inflow rate into the circumbinary disc)
\begin{equation}
    \dot{M}_{\rm BHL, CoM} = \alpha \frac{4\upi G^2 M_{\rm bin}^2 \rho_{\rm CoM}}{\left(c_{\rm s, CoM}^2 + v_{\rm rel, CoM}^2\right)^{3/2}},
\end{equation}
where $M_{\rm bin} = M_{\rm BH, 1} + M_{\rm BH, 2}$ is the total mass of the binary system, and the gas density ($\rho_{\rm CoM}$), sound speed ($c_{\rm s, CoM}$), and the magnitude of relative velocity between the gas and the binary's CoM velocity ($v_{\rm rel, CoM}$) are computed at the CoM position of the binary. Again $\alpha = 25$ is adopted in the binary accretion model. We also cap the total accretion rate by the total Eddington accretion rate,
\begin{equation} \label{eq:bin_acc_total}
    \dot{M}_{\rm bin} \equiv \dot{M}_{\rm BH, 1} + \dot{M}_{\rm BH, 2} = {\rm min}(\dot{M}_{\rm BHL, CoM}, \dot{M}_{\rm Edd, bin}),
\end{equation}
where the total Eddington accretion rate is
\begin{equation}
    \dot{M}_{\rm Edd, bin} = \frac{4 \upi G M_{\rm bin} m_{\rm p}}{\epsilon_{\rm r} \sigma_{\rm T} c}.
\end{equation}

\begin{figure} 
\centering\includegraphics[width=\columnwidth]{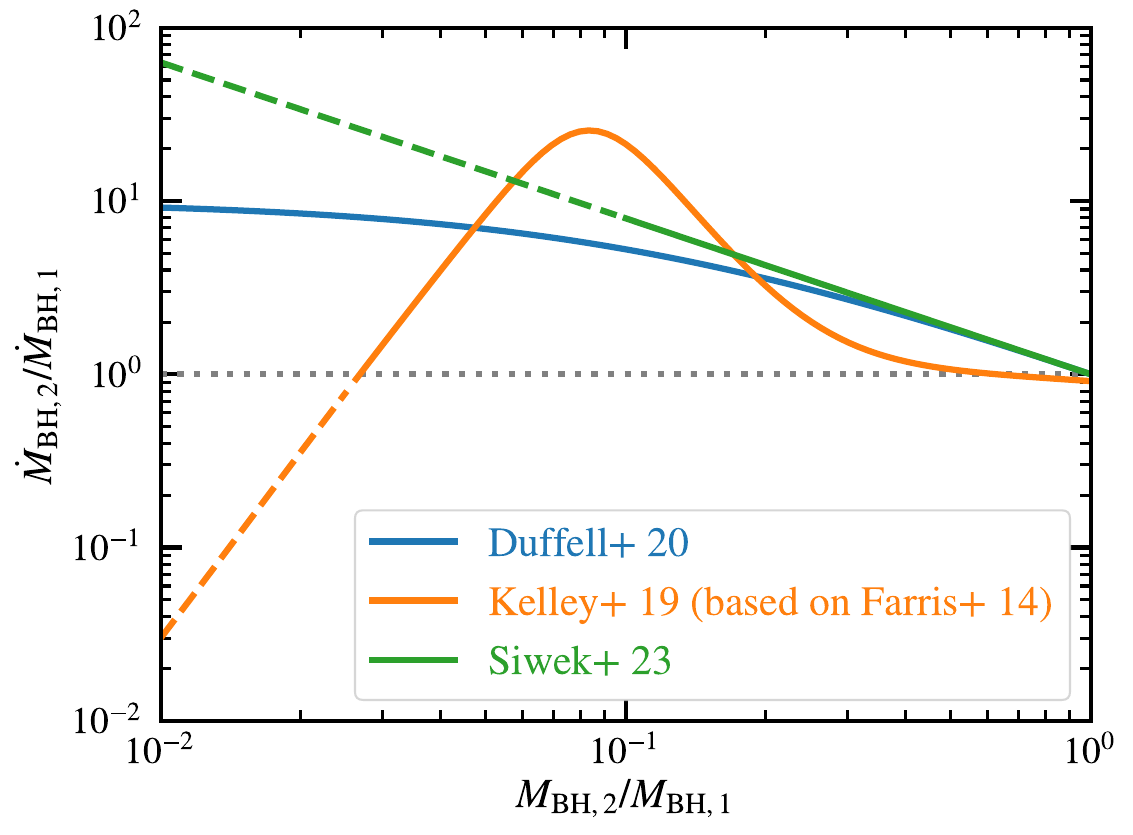}
\caption{Fitting formulae from circumbinary disc simulations for the BH accretion rate ratios as a function of the BH mass ratios. The formulae from \citet{Duffell2020}, \citet{Kelley2019}, and \citet{Siwek2022} are plotted with blue, orange, and green lines respectively. The solid segment marks the actual mass-ratio range which is used for the fitting while the dashed segment plots the fitting formulae beyond the actual fitting range. The horizontal dotted line marks $\dot{M}_{\rm BH, 2}/\dot{M}_{\rm BH, 1} = 1$.}
\label{fig:cbd_fit_eq}
\end{figure}

To further compute the actual mass accretion rate for each BH, we adopt the fitting equations from recent circumbinary disc simulations. In Fig.~\ref{fig:cbd_fit_eq}, we compare the fitting formulae for the BH accretion rate ratio ($\dot{M}_{\rm BH, 2}/\dot{M}_{\rm BH, 1}$) as a function of the BH mass ratio ($q \equiv M_{\rm BH, 2} / M_{\rm BH, 1}$) from \citet{Kelley2019}, \citet{Duffell2020}, and the recent work of \citet{Siwek2022}. Note that the \citet{Kelley2019} formula is based on the data points from the \citet{Farris2014} simulations in the mass ratio range of $0.026 \leq q \leq 1$, but it gives $\dot{M}_{\rm BH, 2}/\dot{M}_{\rm BH, 1} < 1$ for $q \ga 0.6$ which contradicts the conclusion from circumbinary disc simulations. \citet{Duffell2020} provide an updated fitting formula using simulations with mass ratios down to $q \sim 0.01$. The \citet{Siwek2022} formula is based on simulations with BH mass ratio ranging from $q=0.1$ to $q=1$.

Different fitting formulae tend to agree with each other better for larger mass ratios, i.e. $q \ga 0.1$ (especially for recent simulations), the range which we also explore with our simulated BH binaries in this study, whereas they exhibit larger deviations for smaller $q$ (which are less explored in circumbinary disc simulations). The difference among different formulae might originate from the different numerical setups and methods used in these simulations (e.g. the viscosity model, sink prescription); see the discussions in e.g. \citet{Dittmann2021,Siwek2022}. 

In this study, we adopt the \citet{Duffell2020} formula which is fitted from simulations with the broadest mass ratio range and agrees well with other recent simulations such as \citet{Munoz2020} and \citet{Siwek2022},
\begin{equation} \label{eq:bin_acc_ratio}
    \frac{\dot{M}_{\rm BH, 2}}{\dot{M}_{\rm BH, 1}} = \frac{1}{0.1 + 0.9 q}.
\end{equation}
With Eqs (\ref{eq:bin_acc_total}) and (\ref{eq:bin_acc_ratio}), we can determine the mass accretion rate for each BH in the binary. Note that as we cap the total accretion rate by the total Eddington limit in our model, the secondary BH can have super-Eddington accretion (compared with the Eddington limit computed by its own BH mass) if the BH mass ratio is small enough.

We denote the binary accretion model introduced above as `{\bf ketju + binary accretion}' in this paper. Note that, as in the ketju + single accretion model, here the repositioning prescription is not used for BHs with ketju integration and BHs are merged when their separation is less than 6 times the combined Schwarzschild radius.

\subsection{BH feedback}\label{subsec:bh_feedback}

To model the feedback from BHs, in this work, we implement two approaches, i.e. thermal feedback and kinetic feedback. In the thermal feedback, following the previous literature \citep[e.g.][]{DiMatteo2005,Springel2005BH}, we assume that some fraction $\epsilon_{\rm f, th}$ of the radiated luminosity,
\begin{equation}
    \dot{E}_{\rm th} = \epsilon_{\rm f, th} \epsilon_{\rm r} \dot{M}_{\rm BH} c^2,
\end{equation}
is coupled to the surrounding gas thermally and isotropically. We use $\epsilon_{\rm f, th} = 0.02$ which reproduces the observed BH mass--stellar velocity dispersion ($M_{\rm BH}$--$\sigma_{\star}$) relation. At each time step, this thermal feedback energy is added to the internal energy of every gas particle within $h_{\rm BH}$ according to the SPH kernel weights, i.e. the $i$-th gas particle receives the amount of energy
\begin{equation}
    E_{{\rm th},i} = \frac{w_i m_i \dot{E}_{\rm th} \Delta t}{\rho},
\end{equation}
where $m_i$ is the gas particle mass.

Motivated by observations of the AGN driven galactic outflows with velocities ranging from a few thousand ${\rm km}~{\rm s}^{-1}$ to ${\sim} 0.1c$ \citep[see e.g.][for reviews]{Crenshaw2003,Laha2021},  the kinetic feedback mechanism has been introduced into galaxy formation simulations and has been shown to be more efficient than thermal feedback in regulating the BH growth, driving higher-velocity gas outflows, and shock-heating gas compared to the thermal feedback model \citep[e.g.][]{Choi2012,Debuhr2012,Barai2014,Choi2014,Choi2015,Barai2016,Angles-Alcazar2017,Costa2018,Costa2020,Torrey2020}. For the implementation of kinetic feedback in this study, we adopt an approach which is similar to the kinetic mode of the AGN feedback in the Illustris TNG simulations \citep{Weinberger2017}. 

Similar to the thermal feedback approach, we assume that some fraction $\epsilon_{\rm f, kin}$ of the radiated luminosity,
\begin{equation}
    \dot{E}_{\rm kin} = \epsilon_{\rm f, kin} \epsilon_{\rm r} \dot{M}_{\rm BH} c^2,
\end{equation}
is coupled to the gas particles within the BH smoothing length. But unlike the continuous energy feedback in the thermal approach, we add this kinetic feedback energy to the energy reservoir of the BH at each time step, i.e.
\begin{equation}
    E_{\rm kin, res}(t + \Delta t) = E_{\rm kin, res}(t) + \dot{E}_{\rm kin} \Delta t.
\end{equation}
The energy stored in the reservoir is released to the surrounding gas when it reaches the threshold,
\begin{equation}
    E_{\rm thr} = \frac{1}{2} f_{\rm thr} M_{\rm gas}(< h_{\rm BH}) \sigma_{\rm DM}^2,
\end{equation}
where $f_{\rm thr}$ is a free parameter that controls the strength of each kinetic feedback energy pulse, $M_{\rm gas}(< h_{\rm BH})$ is the enclosed gas mass within the BH smoothing length, and $\sigma_{\rm DM}$ is the one-dimensional velocity dispersion computed from the $40$ nearest neighbouring dark matter particles of the BH. By linking the feedback energy threshold to the local dark matter velocity dispersion, we ensure that the wind energy does not significantly exceed the galaxy binding energy, and therefore we will not unbind a large amount of gas. The use of the dark matter velocity dispersion, instead of the gas or stellar velocity dispersions, provides a more numerically robust estimation of the local gravitational potential, as some galaxies might not have enough gas/star particles.

When releasing the energy in the reservoir, each surrounding gas particle receives a fraction of the energy according to the SPH kernel weights, i.e.
\begin{equation}
    E_{{\rm kin}, i} = \frac{w_i m_i E_{\rm kin, res}}{\rho}.
\end{equation}
This amount of energy is coupled to the gas particle in a kinetic form by adding a kick velocity with the magnitude of
\begin{equation}
    v_{{\rm kick}, i} = \sqrt{\frac{2E_{{\rm kin}, i}}{m_i}}.
\end{equation}
In the single BH phase, following \citet{Choi2012}, we set the direction of the kick velocity to be parallel ($50$ per cent probability) or antiparallel ($50$ per cent probability) to the direction of angular momentum $\mathbfit{r}_{i} \times \mathbfit{v}_{i}$, where $\mathbfit{r}_{i}$ and $\mathbfit{v}_{i}$ are the $i$th gas particle's position and velocity vectors with respect to the BH. In the binary BH phase, the kick direction is parallel or anti-parallel (again $50$ per cent probability for each possibility) to the orbital angular momentum direction of the BH binary.

Note that compared to many previous works of kinetic AGN feedback in which the kick velocity is fixed to a constant value \citep[usually in the range of ${\sim} 10^{3}$--$10^{4} ~ {\rm km}~{\rm s}^{-1}$, see e.g.][]{Choi2012,Debuhr2012,Barai2014,Angles-Alcazar2017}, the kick velocity in our model varies for different gas particles, and more physically, the overall kick velocity magnitude depends on the local gravitational potential or equivalently the galaxy virial mass.

Our kinetic model has two free parameters, $\epsilon_{\rm f, kin}$ and $f_{\rm thr}$. As shown in Appendix \ref{ap:kin_fb}, the final BH mass of the galaxy merger remnant is more sensitive to the feedback efficiency parameter $\epsilon_{\rm f, kin}$, i.e. with a larger (smaller) $\epsilon_{\rm f, kin}$, the final BH mass is lower (higher). Therefore, we can use the observed $M_{\rm BH}$--$\sigma_{\star}$ relation to calibrate the value for $\epsilon_{\rm f, kin}$, which leads to $\epsilon_{\rm f, kin} = 0.008$ for this study. The energy reservoir threshold parameter $f_{\rm thr}$ controls the magnitudes of the kick velocity in each feedback event. As shown in Appendix \ref{ap:kin_fb}, with the fiducial value adopted for the simulations in this study $f_{\rm thr} = 20$, the maximum kick velocities can reach up to values of ${\sim} 5000$--$10000$ ${\rm km}~{\rm s}^{-1}$, which are the typical values of the observed outflow velocity \citep[e.g.][]{Crenshaw2003,Laha2021}. The details of the tests for our kinetic feedback model are summarised in Appendix \ref{ap:kin_fb}.

Similar to some previous works which studied thermal and kinetic feedback models in simulations of isolated and merging galaxies \citep[e.g.][]{Barai2014,Barai2016}, in this work, we use either pure thermal or pure kinetic feedback model. This approach allows us to compare the impact of different AGN feedback models on the results.

\section{Simulations}\label{sec:sim}

\subsection{Initial conditions}\label{subsec:sim_ic}

\begin{table*}
\begin{threeparttable}
\caption{Single progenitor galaxies}
\label{tab:prog_info}
\begin{tabular}{lccccccccccc}
\hline
Name & $V_{\rm vir}$ & $R_{\rm vir}$ & $r_{\rm d}$ & $M_{\rm tot}$ & $M_{\rm DM}$ & $M_{\star}$ & $M_{\rm gas}$ & $M_{\rm BH}$ & $N_{\rm DM}$ & $N_{\star}$ & $N_{\rm gas}$\\
& $[{\rm km}~{\rm s}^{-1}]$ & $[{\rm kpc}]$ & $[{\rm kpc}]$ & $[10^{10}~{\rm M}_{\sun}]$ & $[10^{10}~{\rm M}_{\sun}]$ & $[10^{10}~{\rm M}_{\sun}]$ & $[10^{10}~{\rm M}_{\sun}]$ & $[10^{5}~{\rm M}_{\sun}]$ & $[10^5]$ & $[10^5]$ & $[10^5]$ \\
\hline
D1 & $200$ & $282$ & $4.32$ & $262$ & $248$ & $12.2$ & $2.15$ & $1$ & $16.00$ & $12.18$ & $2.15$\\
D2 & $250$ & $352$ & $5.40$ & $512$ & $484$ & $23.8$ & $4.20$ & $1$ & $32.00$ & $23.80$ & $4.20$\\
\hline
\end{tabular}
\begin{tablenotes}[flushleft]
 \footnotesize
 \item {\it Note}. From the left, the name of progenitor galaxy, the virial velocity $V_{\rm vir}$, the virial radius $R_{\rm vir}$, the disc scale length $r_{\rm d}$, the total mass $M_{\rm tot}$, the dark matter mass $M_{\rm DM}$, the stellar mass $M_{\star}$ (including both disc and bulge stars), the gas mass $M_{\rm gas}$, the initial BH mass $M_{\rm BH}$, the dark matter particle number $N_{\rm DM}$, the star particle number $N_{\star}$, and the gas particle number $N_{\rm gas}$. 
\end{tablenotes}
\end{threeparttable}
\end{table*}

\begin{table*}
\begin{threeparttable}
\caption{Galaxy merger samples}
\label{tab:merger_info}
\begin{tabular}{lccccccc}
\hline
Name & Primary & Secondary & Orbit & Galaxy mass & $N_{\rm DM, tot}$ & $N_{\star,{\rm tot}}$ & $N_{\rm gas, tot}$\\
& progenitor & progenitor & geometry & ratio & $[10^5]$ & $[10^5]$ & $[10^5]$ \\
\hline
DD-11-G0 & D1 & D1 & G0 & $1:1$ & $32.00$ & $24.36$ & $4.30$ \\
DD-11-G5 & D1 & D1 & G5 & $1:1$ & $32.00$ & $24.36$ & $4.30$ \\
DD-11-G15 & D1 & D1 & G15 & $1:1$ & $32.00$ & $24.36$ & $4.30$ \\
DD-21-G0 & D2 & D1 & G0 & $2:1$ & $48.00$ & $35.98$ & $6.35$ \\
DD-21-G5 & D2 & D1 & G5 & $2:1$ & $48.00$ & $35.98$ & $6.35$ \\
DD-21-G15 & D2 & D1 & G15 & $2:1$ & $48.00$ & $35.98$ & $6.35$ \\
\hline
\end{tabular}
\begin{tablenotes}[flushleft]
 \footnotesize
 \item {\it Note}. The three rightmost columns give the total dark matter particle number $N_{\rm DM, tot}$ (including both primary and secondary progenitors), the total star particle number $N_{\star, {\rm tot}}$, and the total gas particle number $N_{\rm gas, tot}$. 
\end{tablenotes}
\end{threeparttable}
\end{table*}

To set up the initial conditions for our galaxy mergers, we first follow the method in \citet{Springel2005BH} to generate single progenitor galaxies, then set up the orbit and orientation of the two progenitor galaxies. In all initial conditions, the gas and star particle mass is $m_{\rm gas} = m_{\star} = 10^{5} ~ {\rm M}_{\sun}$, and the dark matter particle mass is $m_{\rm DM} \approx 1.6 \times 10^{6} ~ {\rm M}_{\sun}$. The initial BH particle mass is set to the same value as the other baryonic particles, mimicking the BH seeding in cosmological simulations and this places the progenitor galaxies below the observed $M_{\rm BH}$--$\sigma_{\star}$ relation. The cosmic time of the initial condition is set to $10.7~{\rm Gyr}$.

We generate two disc progenitor galaxies, D1 and D2, with different total masses. Each disc progenitor consists of a dark matter halo with a \citet{Hernquist1990} density profile, an exponential gas disc, an exponential stellar disc, a Hernquist stellar bulge, and a BH. Except for the input virial velocity ($V_{\rm vir}$) which is used to determine the total mass, all other input parameters are the same when generating D1 and D2. Specifically, the input Navarro--Frenk--White \citep*[NFW,][]{Navarro1996} profile-equivalent halo concentration is $c = 9$. The spin parameter is $\lambda = 0.033$, the total disc (i.e. gas + stellar) mass fraction is $m_{\rm d} = 0.041$, the gas mass fraction in the disc is $f_{\rm gas} = 0.2$, the fractional disc angular momentum is $j_{\rm d} = m_{\rm d}$, and the vertical height of the stellar disc is set to be $z_{\rm d} = 0.2 r_{\rm d}$ with $r_{\rm d}$ being the disc scale length. We assume here that the gas disc and the stellar disc have the same scale length. The stellar bulge mass fraction is $m_{\rm b} = m_{\rm d}/3 = 0.01367$, and the bulge scale length is set to $r_{\rm b} = 0.2 r_{\rm d}$. Furthermore, we follow \citet{Lahen2018} to initialize the stellar ages and stellar/gaseous metallicities. To set up the stellar ages for bulge (disc) star particles, we assume an exponentially (linearly) decaying star formation rate. To initialize the metal abundance of disc stars, bulge stars, and gas particles, we adopt a metallicity gradient of $k = 0.0585~{\rm dex}~{\rm kpc}^{-1}$, a scale radius $r_{\rm s} = 3~{\rm kpc}$, and the initial abundances at $r_{\rm s}$ as given in Table 2 of \citet{Lahen2018}. The helium abundance is set according to the observed helium-metallicity relation, $Y = 2.1 Z + 0.24$, with $Y$ and $Z$ being the helium and total metal mass fractions respectively \citep{Jimenez2003,Casagrande2007}.

For D1, the virial velocity is $V_{\rm vir} = 200 ~ {\rm km}~{\rm s}^{-1}$, and thus the total mass is $M_{\rm tot} = 2.62 \times 10^{12}~{\rm M}_{\sun}$ and the virial radius\footnote{The virial radius is defined as the radius within which the mean density is $200$ times the critical density.} is $R_{\rm vir} = 282~{\rm kpc}$. The disc scale length for D1 is $r_{\rm d} = 4.32~{\rm kpc}$. The total mass of galaxy D2 is roughly twice the total mass of D1. For D2, the input virial velocity is $V_{\rm vir} = 250~{\rm km}~{\rm s}^{-1}$, resulting in a total mass of $M_{\rm tot} = 5.12 \times 10^{12}~{\rm M}_{\sun}$ and a virial radius of $R_{\rm vir} = 352~{\rm kpc}$. The D2 disc scale length is $r_{\rm d} = 5.40~{\rm kpc}$. See Table~\ref{tab:prog_info} for more detailed information on these progenitors (e.g. the masses and particle numbers for the different components).  The reason that we have relatively more massive disc galaxy progenitors here compared to previous works \citep[e.g.][]{Johansson2009ApJ,Johansson2009ApJL} is to ensure that the BH seeds can grow to be massive enough before they enter the binary phase, fulfilling the BH-to-stellar particle mass ratio requirements for the ketju integration (for details see Section \ref{subsec:sim_details}).

The progenitor galaxies described above are further used to generate the initial condition for the galaxy mergers. We assume that the two galaxies approach each other on a parabolic orbit with an initial separation of $d_{\rm sep, ini} = (R_{\rm vir, p} + R_{\rm vir, s}) / 2$ and a pericentre distance of $d_{\rm peri, ini} = r_{\rm d, p} + r_{\rm d, s}$. Here, the subscripts of `p' and `s' represent `primary' (more massive) galaxy and `secondary' (less massive) galaxy respectively. The assumption of parabolic orbit is motivated by the work of \citet{Khochfar2006}, which shows that half of the galaxy merger orbits are parabolic or very close to parabolic in cosmological simulations.

In order to have a diverse merger sample, we set up initial conditions with different mass ratios and orbital geometries. Specifically, for the initial disc orientations, we adopt the geometries of G0, G5, and G15 from \citet{Naab2003}, of which the inclination of the disc with respect to the orbit plane $i$ and the argument of pericentre $\omega$ for primary (`p') and secondary (`s') disc galaxies are as follows:

G0: $i_{\rm p} = \omega_{\rm p} = i_{\rm s} = \omega_{\rm s} = 0^\circ$,

G5: $i_{\rm p} = -109^\circ$, $\omega_{\rm p} = -60^\circ$, $i_{\rm s} = 180^\circ$, $\omega_{\rm s} = 0^\circ$,

G15: $i_{\rm p} = -109^\circ$, $\omega_{\rm p} = 60^\circ$, $i_{\rm s} = 71^\circ$, $\omega_{\rm s} = -30^\circ$.

The merger samples that we consider in this study are summarised in Table \ref{tab:merger_info}. We name the galaxy mergers using the convention of `progenitor galaxy types-galaxy mass ratio-orbit geometry', for example, `DD-21-G15' denotes the merger of two disc galaxies with an initial mass ratio $2:1$ and an initial G15 orbit. 

\subsection{Simulation details}\label{subsec:sim_details}

All simulations start at the cosmic time of $t_{0} = 10.7~{\rm Gyr}$ and are evolved for $3~{\rm Gyr}$, with the final snapshots reaching the current age of the Universe (e.g. ${\sim} 13.7~{\rm Gyr}$). All BHs merge between $1.4$ and $2~{\rm Gyr}$ after the start of the simulations. Each galaxy merger configuration consists of six runs: `ketju + binary accretion', `ketju + single accretion', and `gadget', which differ in the BH accretion model as described in Section~\ref{sec:bh_model}, and each of them has been run with both a pure thermal and a pure kinetic feedback model.

Note that a high enough BH-to-stellar particle mass ratio (e.g. at least a few hundred) is required to give sufficiently converged binary dynamics \citep[][]{Rantala2017}. Therefore, in ketju runs, we only switch on the ketju integration for a BH when the mass of this BH reaches $M_{\rm BH} \geq 3 \times 10^{7}~{\rm M}_{\sun}$, i.e. reaching a BH-to-stellar particle mass ratio $M_{\rm BH}/m_{\star} \geq 300$, which always happens before the binary phase in all runs. When two BHs enter the binary phase, typically this ratio can reach $M_{\rm BH}/m_{\star} \ga 400$. Before switching on the ketju integration, these runs are the same as the gadget runs. Furthermore, in the ketju runs, when two BHs have merged, there is no further BH merger in the remaining simulation time, therefore, to reduce computational time, we switch off the ketju integration, and move back to the gadget mode but without repositioning the BH at each time-step\footnote{When the simulations return back to the gadget mode, repositioning is no longer used because the merged BH has now become massive enough compared to the star particles (e.g. $M_{\rm BH}/m_\star \ga 1000$). Note that a similar approach has been adopted in the EAGLE simulation, i.e. only BHs with masses less than $ 100m_{\rm gas}$ are repositioned to the minimum of the halo gravitational potential \citep{Schaye2015}.}. To summarise, the ketju runs have three stages, i.e. they are first run with the gadget mode (with BH repositioning) when the BH mass is below $3 \times 10^{7}~{\rm M}_{\sun}$, then we switch on the ketju integration until the two BHs merge, and finally move back to the gadget mode (without BH repositioning).

The softening lengths for dark matter, gas, stellar, and BH particles are $\epsilon_{\rm DM} = 100~{\rm pc}$, $\epsilon_{\rm gas} = 20~{\rm pc}$, $\epsilon_{\star} = 5~{\rm pc}$, and $\epsilon_{\rm BH} = 5~{\rm pc}$, respectively. Note that $\epsilon_{\star}$ and $\epsilon_{\rm BH}$ in our simulations are a factor of a few smaller than those in previous works which used the same galaxy formation subgrid model and similar mass resolutions. For example, the stellar mass resolution and softening length are $10^{5}~{\rm M}_{\sun}$ and $20~{\rm pc}$ in \citet{Eisenreich2017}, $1.3 \times 10^{5}~{\rm M}_{\sun}$ and $44~{\rm pc}$ in \citet{Nunez2017}, and $3 \times 10^{5}~{\rm M}_{\sun}$ and $30~{\rm pc}$ in \citet{Mannerkoski2022}. The motivation to adopt smaller $\epsilon_{\star}$ and $\epsilon_{\rm BH}$ here are (i) to resolve the central stellar distribution down to smaller radii, and (ii) to ensure that even for galaxy mergers with the strongest starburst during the merger (i.e. equal-mass mergers with the G0 orbit), the number of star particles within the ketju region radius ($r_{\rm ketju} = 3 \epsilon_{\star}$) is still computationally feasible ($\la 10^4$ particles) with the current \ketju code. As discussed in \citet{Ludlow2020}, based on the EAGLE galaxy formation model \citep{Schaye2015}, the optimal/calibrated parameters for the subgrid model in galaxy formation simulations might depend on the adopted softening lengths. However, as detailed in Section \ref{sec:res}, our simulations reproduce very well the observed scaling relations, thus suggesting that the softening lengths adopted in our simulations are still within a physically motivated range for our adopted galaxy formation subgrid model.

In all simulations, we set the gadget integrator error tolerance parameter to $\eta = 0.02$, and the tree force accuracy parameter to $\alpha_{\rm force} = 0.005$. The ketju integrator Gragg--Bulirsch--Stoer \citep[GBS,][]{Gragg1965,Bulirsch1966} accuracy tolerance parameter is set to $\eta_{\rm GBS} = 10^{-8}$. We save snapshots every $50$ Myr.

\section{Results}\label{sec:res}

\subsection{BH growth history}\label{subsec:bh_growth_his}

\begin{figure*} 
\centering\includegraphics[width=450pt]{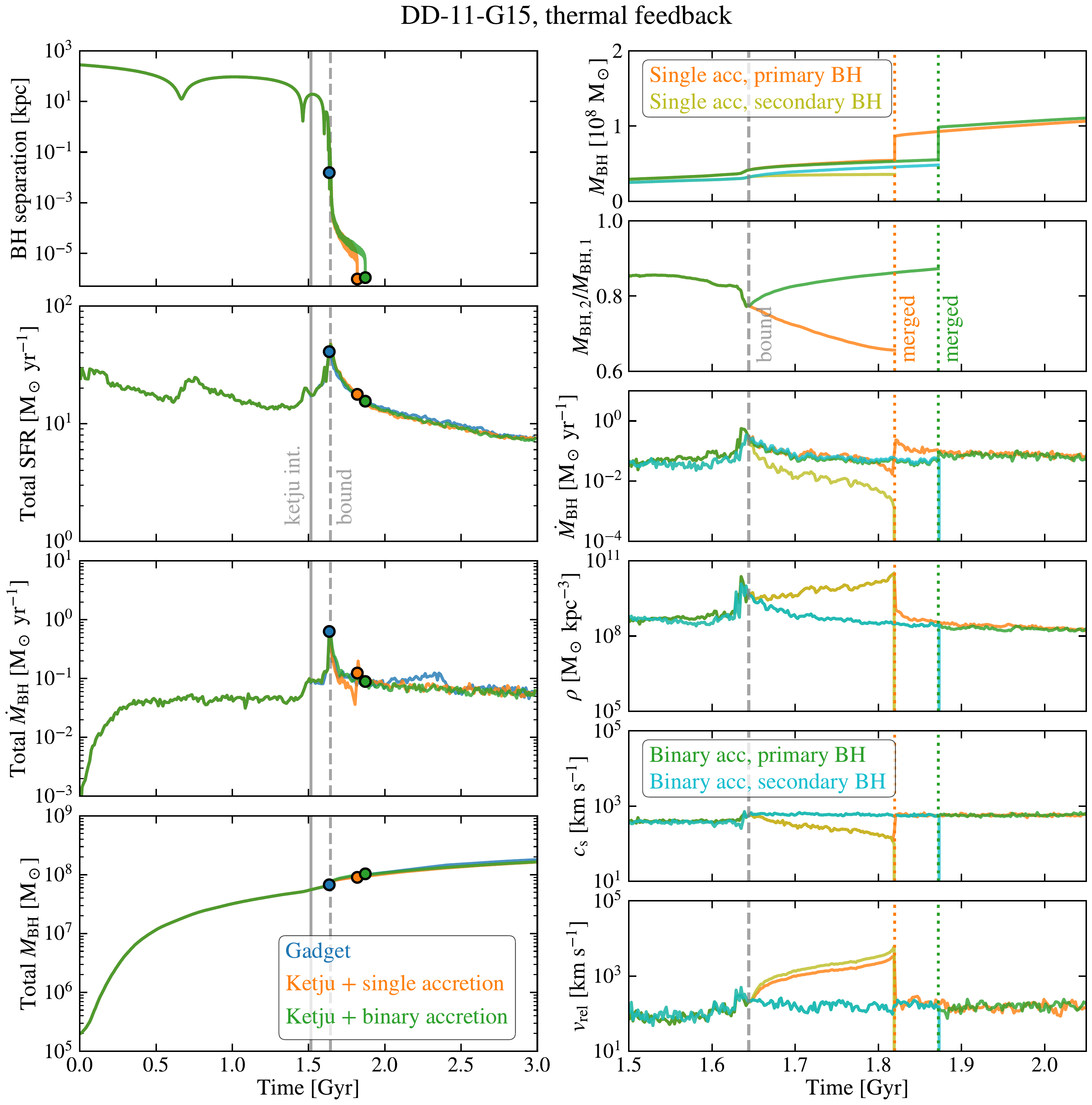}
\caption{The BH growth history in the DD-11-G15 run with thermal feedback. {\it Left-hand panel:} from top to bottom, the time-evolution of the BH separation, the total SFR, the total BH accretion rate, and the total BH mass from the gadget (blue), ketju + single accretion (orange), and ketju + binary accretion (green) runs are plotted. The vertical solid lines mark the onset of the ketju integration for ketju runs, while the vertical dashed lines denote the formation of a bound BH binary in the ketju runs. The filled circles mark the BH merger events. Note that the total SFR and the total BH accretion rate are averaged over 10 Myr. {\it Right-hand panel:} The evolution of the BH mass, the BH mass ratio, the BH accretion rate, the gas density, the sound speed, and the BH-gas relative velocity (from top to bottom) for each BH during the BH binary phase in the ketju runs. The primary and secondary BHs are shown with orange and olive (green and cyan) colours for the ketju + single accretion (ketju + binary accretion) run. The grey vertical dashed lines mark the formation of a bound BH binary, whereas the orange/green vertical dotted lines mark the merging of the BHs in the ketju + single accretion/ketju + binary accretion run. Note that the BH accretion rate, the gas density, the sound speed, and the BH-gas relative velocity are averaged over 2 Myr.}
\label{fig:DD_11_G15_growth}
\end{figure*}

\begin{figure*} 
\centering\includegraphics[width=450pt]{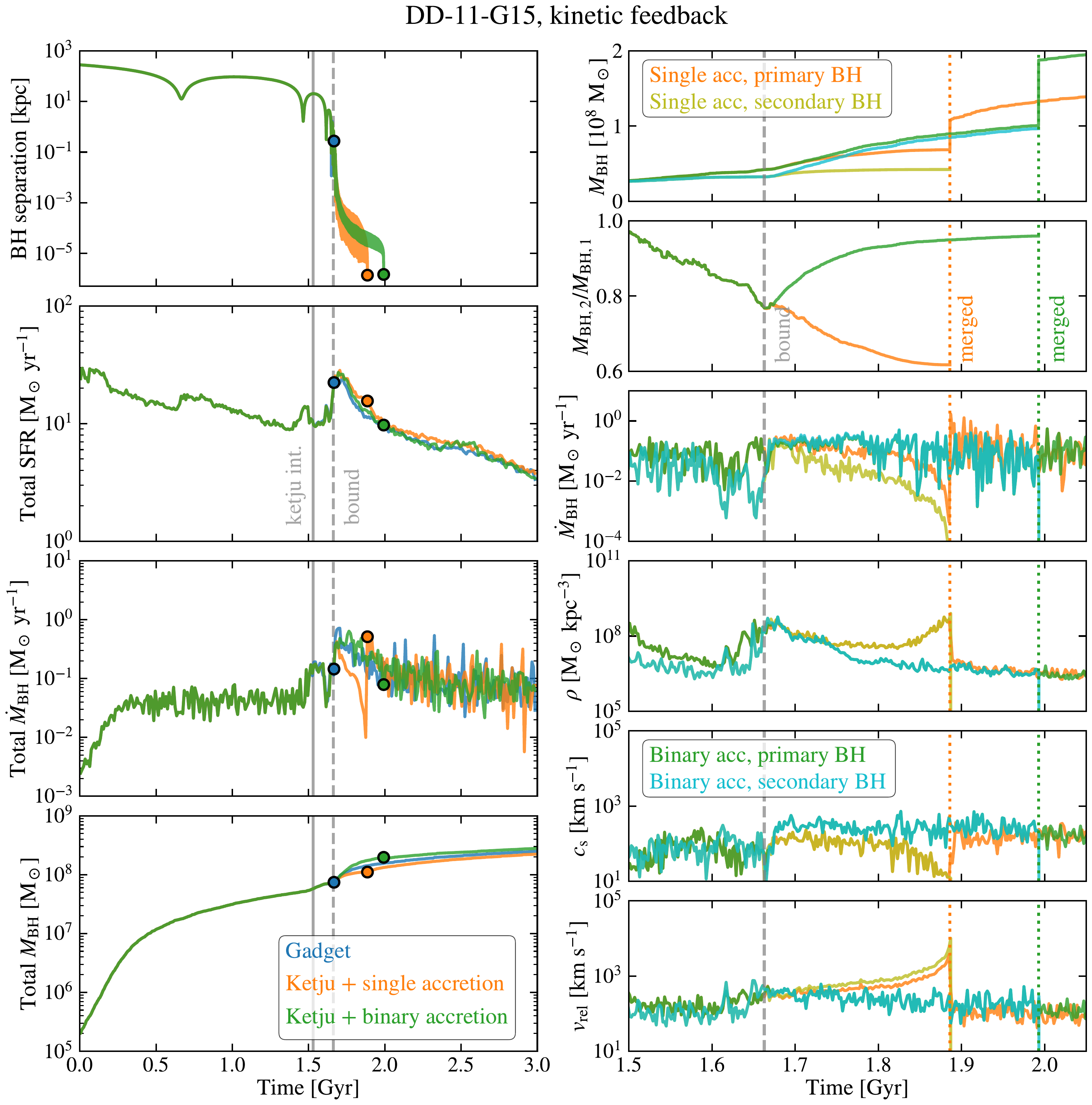}
\caption{Same as Fig.~\ref{fig:DD_11_G15_growth}, but for the BH growth history in the DD-11-G15 run with kinetic feedback.}
\label{fig:DD_11_G15_growth_kin_FB}
\end{figure*}

We start by comparing the BH growth history in merger runs with different BH accretion models. In Fig.~\ref{fig:DD_11_G15_growth}, we use the DD-11-G15 run with thermal feedback as an example to illustrate the behaviour of the BH accretion in the gadget (blue lines), ketju + single accretion (orange lines), and ketju + binary accretion (green lines) models.

In the left panels, we show the time-evolution of the BH separation, the total star formation rate (SFR) in the galaxies, the total BH accretion rate, and the total BH mass in the different runs. Two BHs start from an initial separation of $282~{\rm kpc}$, and evolve exactly in the same way in all three runs until $t \sim 1.5$ Gyr when the mass of a BH in the ketju runs increases to the threshold of $3 \times 10^{7}~{\rm M}_{\sun}$ (marked by the vertical solid lines). Once a BH in a ketju run reaches this mass threshold, it is evolved by the ketju integrator, which computes the non-softened gravity between the BH and star particles in the regularised region. 

After two pericentre passages, the two BHs rapidly sink to the very centre of the galaxy remnant due to dynamical friction. At the same time, the tidal response drives the gas to the centre, and a starburst is triggered, with the total SFR increasing to ${\sim} 60~{\rm M}_{\sun}~{\rm yr}^{-1}$. The BH accretion rates also increase significantly, approaching ${\sim} 1~{\rm M}_{\sun}~{\rm yr}^{-1}$. The two BHs merge at $t \sim 1.65~{\rm Gyr}$ in the gadget run (marked by the blue filled circle) with a final separation of ${\sim} 10~{\rm pc}$, while they become a bound binary roughly at the same time (marked by the vertical dashed lines) in the ketju runs. Once the BHs enter the binary phase in the ketju runs, due to the different BH accretion models, we start to observe differences in the BH evolution between the ketju + single accretion and ketju + binary accretion runs. In the ketju runs, the BH binary subsequently shrinks by interacting with stars and ejecting them in three-body slingshots until the GW emission takes over. BHs in the ketju runs merge roughly ${\sim} 200~{\rm Myr}$ after they become a bound binary at the final separation of ${\sim} 10^{-3}~{\rm pc}$. Compared to the gadget run, the \ketju code enables us to resolve the three-stage BH merging process which is not resolved in traditional galaxy formation simulations.

Overall, all three runs show relatively similar BH growth and global star formation histories as they adopt similar BHL accretion, thermal AGN feedback and subgrid models. However, we do notice that during the BH binary phase, the ketju + binary accretion and ketju + single accretion runs exhibit different BH accretion behaviours, i.e. the total BH accretion rate in the ketju + single accretion run becomes increasingly lower and when the BHs merge there is an abrupt sharp increase in the accretion rate. In contrast, the BH accretion rate is smoother in the ketju + binary accretion run. To see this more clearly, we zoom in on the BH accretion evolution during the binary phase and plot it in the right panels of Fig.~\ref{fig:DD_11_G15_growth}.

In the right panels, we plot the mass, the mass ratio, the accretion rate, the gas density, the sound speed, and the gas-BH relative velocity for each BH in both the ketju + single accretion and ketju + binary accretion runs. Before the BHs enter the binary phase (marked by the vertical dashed lines), these two runs are exactly the same, and the BH mass ratio is decreasing to ${\sim} 0.8$ (as the primary BH accretes more gas according to the BHL formula). After the BHs become a bound binary, in the ketju + binary accretion run we see that the secondary BH has a higher gas accretion rate during the whole binary phase, and therefore the evolutionary trend of the BH mass ratio reverses and approaches instead unity. This is caused by the preferential accretion of the secondary BH in our newly introduced binary accretion model. In contrast, in the ketju + single accretion run, the BHL formula is used for each BH, and the primary BH always has a higher accretion rate since the BH accretion rate is proportional to $M_{\rm BH}^2$. Just before the two BHs merge (marked by the vertical dotted lines), the BH mass ratio in the ketju + binary accretion run is ${\sim} 0.9$, while in the ketju + single accretion run it is ${\sim} 0.65$. This clearly demonstrates that the circumbinary disc subgrid model tends to equalise the masses of the BHs in a binary system. 

Apart from the preferential accretion, we can observe another significant difference between the two ketju runs. As the BHs become tightly bound in the binary phase, their inspiral velocities keep increasing. In the ketju + single accretion run, the gas-BH relative velocity is still computed using the velocity of each BH, and as a result $v_{\rm rel}$ increases from a few hundred to ${\sim} 10^{4}~{\rm km}~{\rm s}^{-1}$. Note also that $v_{\rm rel}$ is larger for the secondary BH. This extremely high relative velocity in the denominator of the BHL formula significantly suppresses the BH accretion rate and thus the AGN feedback strength, and as a result, gas accumulates in the vicinity of the BH. When the two BHs merge, $v_{\rm rel}$ abruptly drops from ${\sim} 10^{4}~{\rm km}~{\rm s}^{-1}$ to ${\sim} 10^{2}~{\rm km}~{\rm s}^{-1}$, and as a consequence the BH accretion rate computed from the BHL formula jumps by roughly three orders of magnitude (note that the plotted jumps in Fig.~\ref{fig:DD_11_G15_growth} are less significant as the accretion rates have been smoothed over $10~{\rm Myr}$ and $2~{\rm Myr}$, in the left and right panels, respectively). The resulting strong AGN feedback subsequently pushes the accumulated gas away, leading to step function-like behaviours in many quantities plotted in the right panels. This flaw of the ketju + single accretion model originates from the fact that although the BH binary dynamics is resolved by the ketju integrator, the gas resolution is still limited in current galactic-scale galaxy formation simulations. In contrast, in the ketju + binary accretion run, as the gas-BH relative velocity is computed using the CoM velocity of the binary when estimating the total accretion rate of the binary system, the plotted quantities evolve more smoothly.

In Fig.~\ref{fig:DD_11_G15_growth_kin_FB}, we plot the BH growth history of the DD-11-G15 merger with pure kinetic AGN feedback. Compared to the thermal feedback run, we notice that the BH accretion in the kinetic run shows more variability, the gas density at the location of the BH is lower, and the SFR is also lower (i.e. the peaked total SFR here is ${\sim} 30~{\rm M}_{\sun}~{\rm yr}^{-1}$ compared to ${\sim} 60~{\rm M}_{\sun}~{\rm yr}^{-1}$ in the thermal case), as the kinetic feedback is more effective in removing gas from the central region and thus reducing the star formation, in agreement with previous studies \citep[e.g.][]{Choi2012,Barai2014,Choi2014,Choi2015,Barai2016}. After the BHs form a binary in the ketju runs, the differences in the evolution of the SFR, the BH accretion rate, and the BH mass among the three runs become larger compared to the thermal feedback case, which should originate from the higher stochasticity of the pulsed kinetic feedback compared to the continuous thermal feedback. Nevertheless, overall the behaviour of the BH accretion in the kinetic feedback runs is qualitatively similar to accretion in the thermal feedback, suggesting that our binary accretion model works in both the thermal and kinetic feedback mechanisms. Finally, note that although here we only use the DD-11-G15 merger as an example to detail the BH growth history, similar results can also be found for the other mergers.

In this subsection, we have demonstrated that if the BHL single accretion model is directly applied to BHs in the resolved binary phase (i.e. the ketju + single accretion model), then the total BH accretion rate is suppressed due to the high BH inspiral velocity and the unresolved small-scale gas velocity. There are also unphysical abrupt jumps in the accretion-related quantities, and the evolution of the BH mass ratio is also in disagreement with the predictions from circumbinary disc simulations. Given these flaws in the ketju + single accretion model, in the following subsections, we will focus on the ketju + binary accretion runs which model the accretion behaviour of binary BHs in a more physically reasonable approach. Unless otherwise specified, the ketju runs in the remaining subsections refer to the ketju + binary accretion runs.

\subsection{Galaxy scaling relations}\label{subsec:scaling_relation}

In this subsection, we compute the scaling relations for the galaxy remnants in all ketju + binary accretion runs and check if the newly introduced BH model in combination with the other included galaxy formation subgrid models can reproduce the observed scaling relations.

\begin{figure} 
\centering\includegraphics[width=\columnwidth]{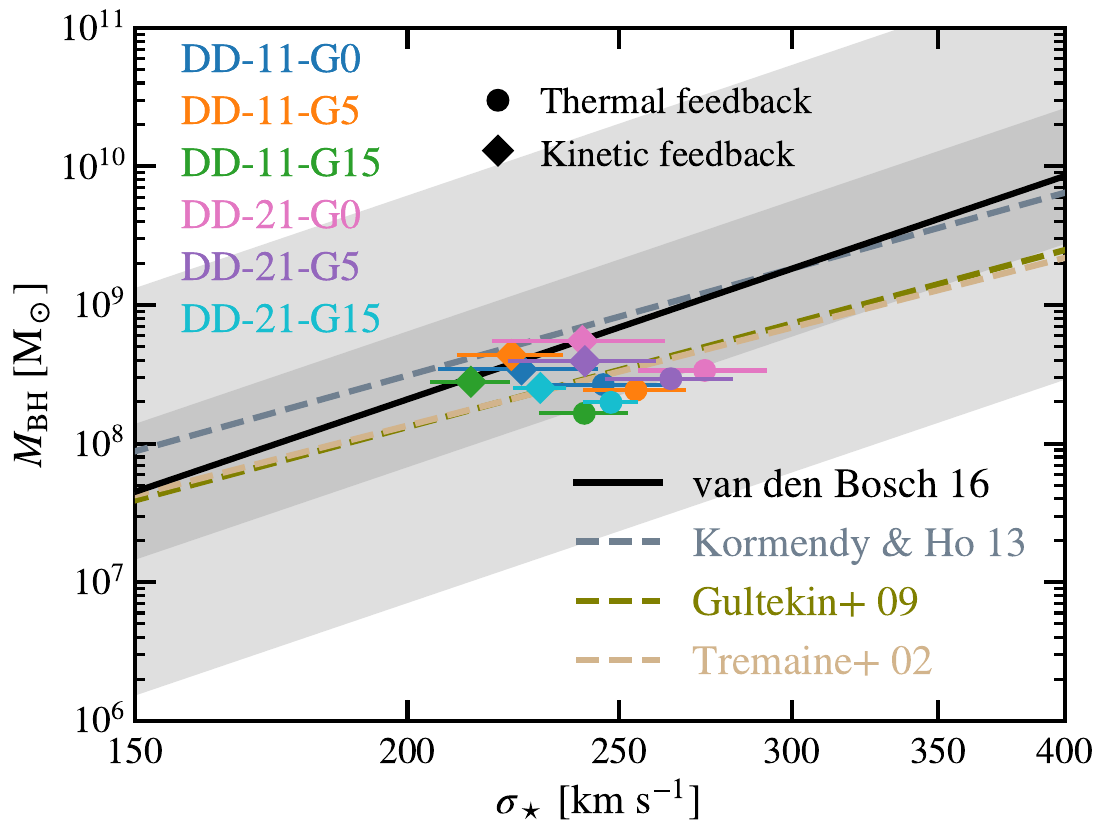}
\caption{The $M_{\rm BH}$--$\sigma_{\star}$ relation for galaxy merger remnants in the ketju + binary accretion runs. The filled circles (diamonds) represent the galaxy remnants of different galaxy mergers marked by different colours in the thermal (kinetic) feedback runs. The standard deviation of the stellar velocity dispersion in each simulation is computed from $50$ random projections. The black solid line shows the best-fitting $M_{\rm BH}$--$\sigma_{\star}$ relation from \citet{van_den_Bosch2016} and the grey-shaded regions denote one and three times the intrinsic scatter. The dashed lines with different colours show the best-fitting relations from \citet{Kormendy2013}, \citet{Gultekin2009}, and \citet{Tremaine2002}.}
\label{fig:M_sigma_relation}
\end{figure}

We first look at the $M_{\rm BH}$--$\sigma_{\star}$ relation. To compute the stellar velocity dispersion, we consider all star particles within an aperture radius of $30~{\rm kpc}$ centred on the BH in the final snapshot ($t = 3~{\rm Gyr}$), and project them along $50$ random line-of-sight (LOS) directions. Along each LOS direction, we compute the projected stellar half-mass radius, $R_{\rm e}$, and then calculate the dispersion of the LOS velocities of all star particles within $R_{\rm e}$. The final $\sigma_{\star}$ and its uncertainty are taken from the mean and standard deviation of all $50$ random projections. The simulated $M_{\rm BH}$--$\sigma_{\star}$ relation from all galaxy merger remnants (distinguished by colours) in both the thermal (filled circles) and the kinetic (diamonds) feedback runs is plotted in Fig.~\ref{fig:M_sigma_relation}, in which we over-plot the observed relations from the literature. Although we start with galaxies containing BH seeds which are located below the observed $M_{\rm BH}$--$\sigma_{\star}$ relation, the final galaxy remnants move onto the observed relation, suggesting that our BH accretion and feedback (both thermal and kinetic) models successfully model the co-evolution of BHs and their host galaxies. We note that the final BHs in our kinetic feedback runs are slightly more massive than those in the thermal feedback runs as our kinetic feedback model was calibrated with the \citet{van_den_Bosch2016} relation while the thermal feedback model was originally calibrated using the \citet{Tremaine2002} relation. Compared to the thermal feedback runs, the overall lower $\sigma_{\star}$ in the kinetic feedback runs reflect the fact that the kinetic feedback mechanism is more effective in suppressing the star formation in the centres of the merger remnants. 

\begin{figure} 
\centering\includegraphics[width=\columnwidth]{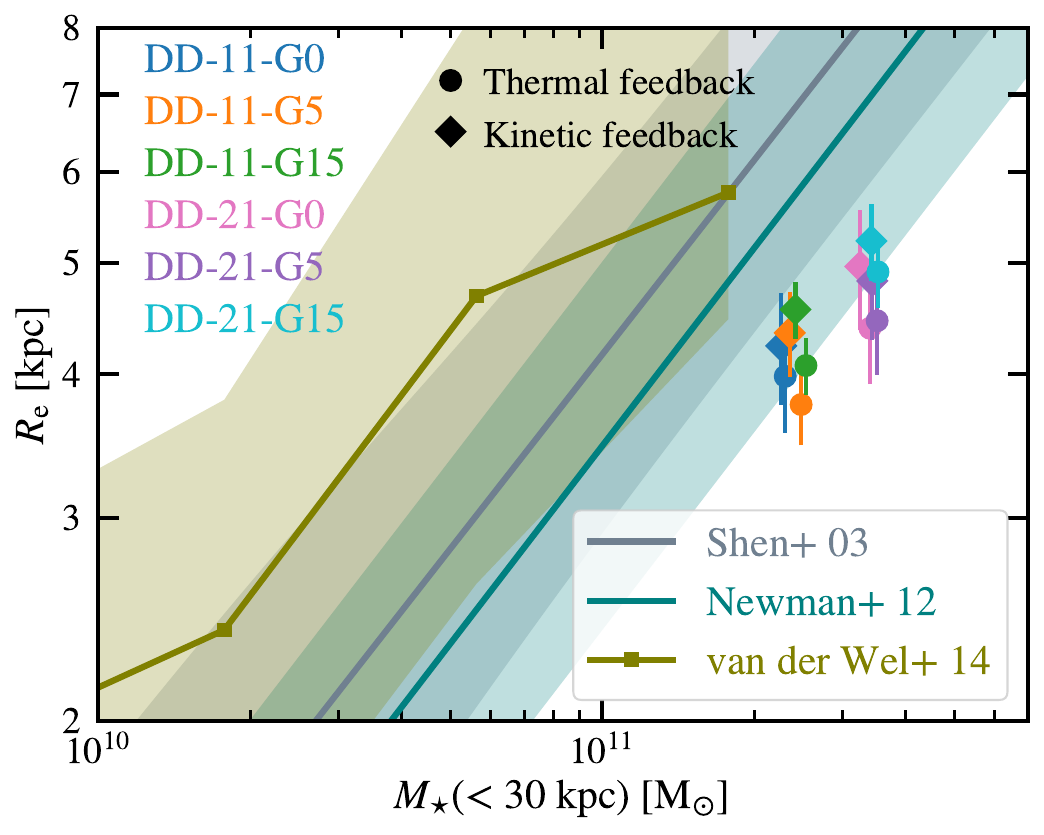}
\caption{The galaxy size--stellar mass relation for the galaxy merger remnants in ketju + binary accretion runs. The filled circles (diamonds) with $y$-errors show the galaxy remnants of different mergers in the thermal (kinetic) feedback runs. Similar to Fig.~\ref{fig:M_sigma_relation}, the error of the projected stellar half-mass radius in each simulation is computed from 50 random projections. The observed relations for early-type galaxies from \citet{Shen2003}, \citet{Newman2012}, and \citet{van_der_Wel2014} are over-plotted for comparison. Note that the redshifts of the galaxies in \citet{Shen2003} are $z \la 0.3$. The \citet{van_der_Wel2014} relation plotted here is the one of the $z=0.25$ bin, and the plotted \citet{Newman2012} relation is the one fitted from $z=0.06$ SDSS early-type galaxies (see their Table 1).}
\label{fig:Mstar_size_relation}
\end{figure}

In Fig.~\ref{fig:Mstar_size_relation}, we compare the simulated galaxy size--stellar mass relation with the observed relations for early-type galaxies from the SDSS \citep{Shen2003,Newman2012} and the 3D-HST+CANDELS \citep{van_der_Wel2014} data. The galaxy size and its uncertainty are computed from the mean and standard deviation of $R_{\rm e}$ along $50$ random LOS directions as described above, and the stellar mass $M_{\star}(< 30~{\rm kpc})$ is the total mass of all star particles within the aperture of $30~{\rm kpc}$. Overall, the sizes of our galaxy remnants are comparable to those of the observed early-type galaxies, but we can also notice that they are systematically somewhat below the best-fitting observed relations. This can be explained by the fact that the galaxy remnants in these idealised merger simulations only experience one major merger, but in the real Universe, galaxies experience additional subsequent minor mergers during their evolution, which can further increase their sizes (e.g. \citealt{Naab2009,Johansson2012}).

\begin{figure} 
\centering\includegraphics[width=\columnwidth]{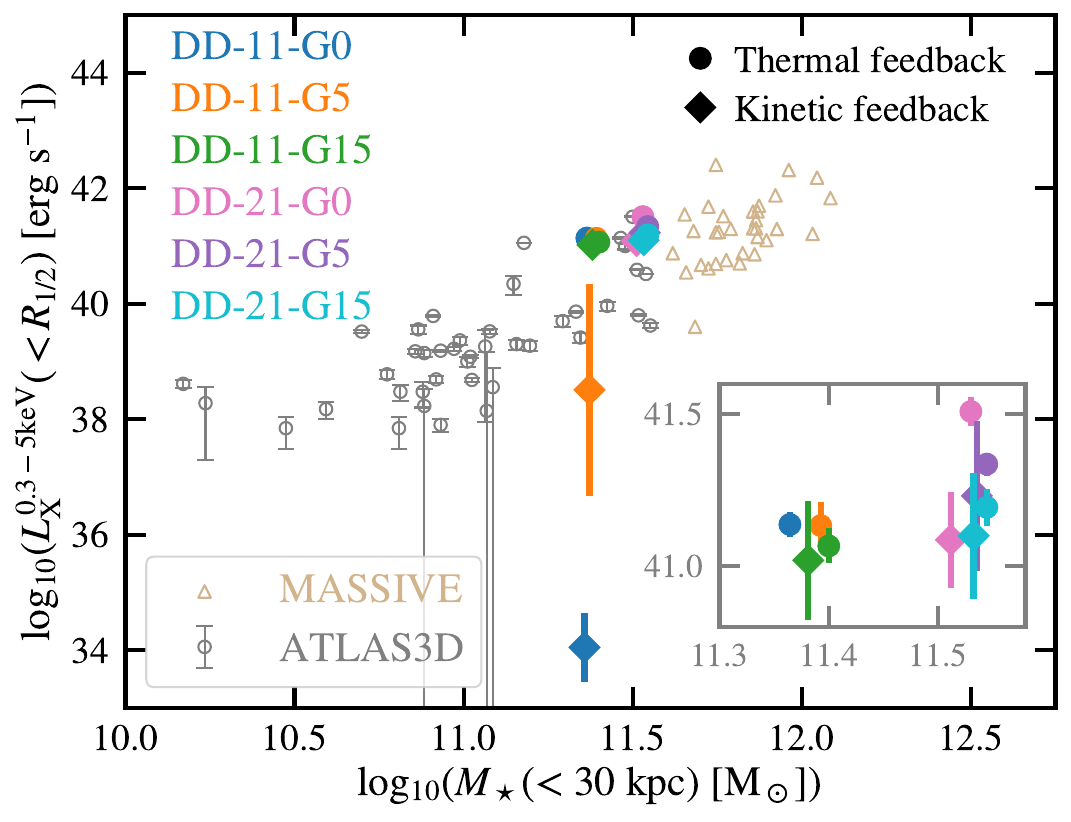}
\caption{The X-ray luminosity from hot and diffuse gas as a function of stellar mass for the galaxy merger remnants in the ketju + binary accretion runs. The filled circles (diamonds) with $y$-errors are from simulations with thermal (kinetic) feedback models. The inset panel shows a zoom of the clustered simulation data points (except the kinetic feedback runs of DD-11-G0 and DD-11-G5). Note that the error bars of some simulation data points cannot be seen if they are smaller than the sizes of those markers. The observations of the MASSIVE and ATLAS3D early-type galaxies \citep{Goulding2016} are plotted with open triangles and circles respectively. Note that the stellar masses for the MASSIVE and ATLAS3D galaxies are estimated from their $K$-band magnitudes, $M_{K}$, which are retrieved from \citet{Ma2014} and \citet{Cappellari2011} respectively, according to the relation of $\log_{10} M_\star = 10.58 - 0.44(M_K + 23)$. See \citet{Cappellari2013} for details.}
\label{fig:LX_Mstar_relation}
\end{figure}

We also compute the X-ray luminosity from hot and diffuse gas for our galaxy remnants and compare the results with observed early-type galaxies from the MASSIVE and ATLAS3D surveys \citep{Goulding2016} in Fig.~\ref{fig:LX_Mstar_relation}. To estimate the X-ray luminosity from hot gas produced by bremsstrahlung and metal-line cooling, we follow the method outlined in \citet{Eisenreich2017}. Using the \textsc{Apec} model (version 3.0.9) \citep{Smith2001} implemented in \textsc{PyAtomDB} \citep{Foster2020}, we first calculate the X-ray luminosity $L_{\rm X}(T, 0.4Z_{\sun})$ and the metallicity gradient $({\rm d} L_{\rm X}/{\rm d}Z)(T)$ in the soft energy range of $0.3$--$5~{\rm keV}$ for a plasma in collisional ionisation equilibrium with a temperature $T$ and metallicity $Z=0.4 Z_{\sun}$ for a range of temperatures between $0.01$ and $85~{\rm keV}$. The solar metal abundance is adopted from \citet{Anders1989}. With this pre-calculated table, for a gas particle with temperature $T_i$ and metallicity $Z_i$ from the simulation, the X-ray luminosity can be estimated as
\begin{equation}
    L_{{\rm X}, i} = N \left[L_{\rm X}(T_i^\prime, 0.4Z_{\sun}) + (Z_i - 0.4 Z_{\sun}) \frac{{\rm d}L_{\rm X}}{{\rm d}Z} \Big|_{T_i^\prime} \right],
\end{equation}
where $T_i^\prime$ is the tabulated temperature which is closest to $T_i$, and the normalisation factor $N$ depends on the mass, mass density, and electron density of the gas particle. The total X-ray luminosity of a galaxy remnant is computed by summing over all considered gas particles. We only consider gas particles for which the distance from the BH are within the 3D stellar half-mass radius ($R_{1/2}$). We further exclude gas particles with densities higher than the star formation density threshold ($2.2 \times 10^{-24}~{\rm g}~{\rm cm}^{-3}$), as such star-forming dense gas is likely to be obscured by the dust. Note that the gas X-ray luminosity of a galaxy fluctuates over time, and in order to have a more robust estimation, we have computed the X-ray luminosities of the last $10$ snapshots (corresponding to the simulation time between $2.5$--$3~{\rm Gyr}$) for each merger remnant, and plot their mean and standard deviation in Fig.~\ref{fig:LX_Mstar_relation}.

From Fig.~\ref{fig:LX_Mstar_relation}, we find that the overall hot gas X-ray luminosities of our galaxy remnants agree well with observations. Compared to the thermal feedback runs, the X-ray luminosities in kinetic feedback runs tend to have larger error bars, reflecting the higher stochasticity of the pulsed kinetic feedback model. We also notice that the mergers of DD-11-G0 and DD-11-G5 with kinetic feedback tend to predict lower X-ray luminosity as these two galaxy mergers (especially DD-11-G0) have higher BH accretion rates and thus stronger AGN feedback during the merger, and the kinetic feedback have removed more gas from the centre for these two mergers thus lowering the gas X-ray luminosity significantly. This is in agreement with \citet{Choi2014} which suggests that kinetic AGN feedback can produce lower gas X-ray luminosity compared to thermal AGN feedback.

To summarise, we have presented three scaling relations that are related to the BHs, the stellar component, and the gas component from our simulations, and have shown that the newly introduced BH model in combination with the subgrid models adopted in our simulations are able to reproduce the observations reasonably well.

\subsection{BH binary orbital evolution}

\begin{figure*}
\centering\includegraphics[width=425pt]{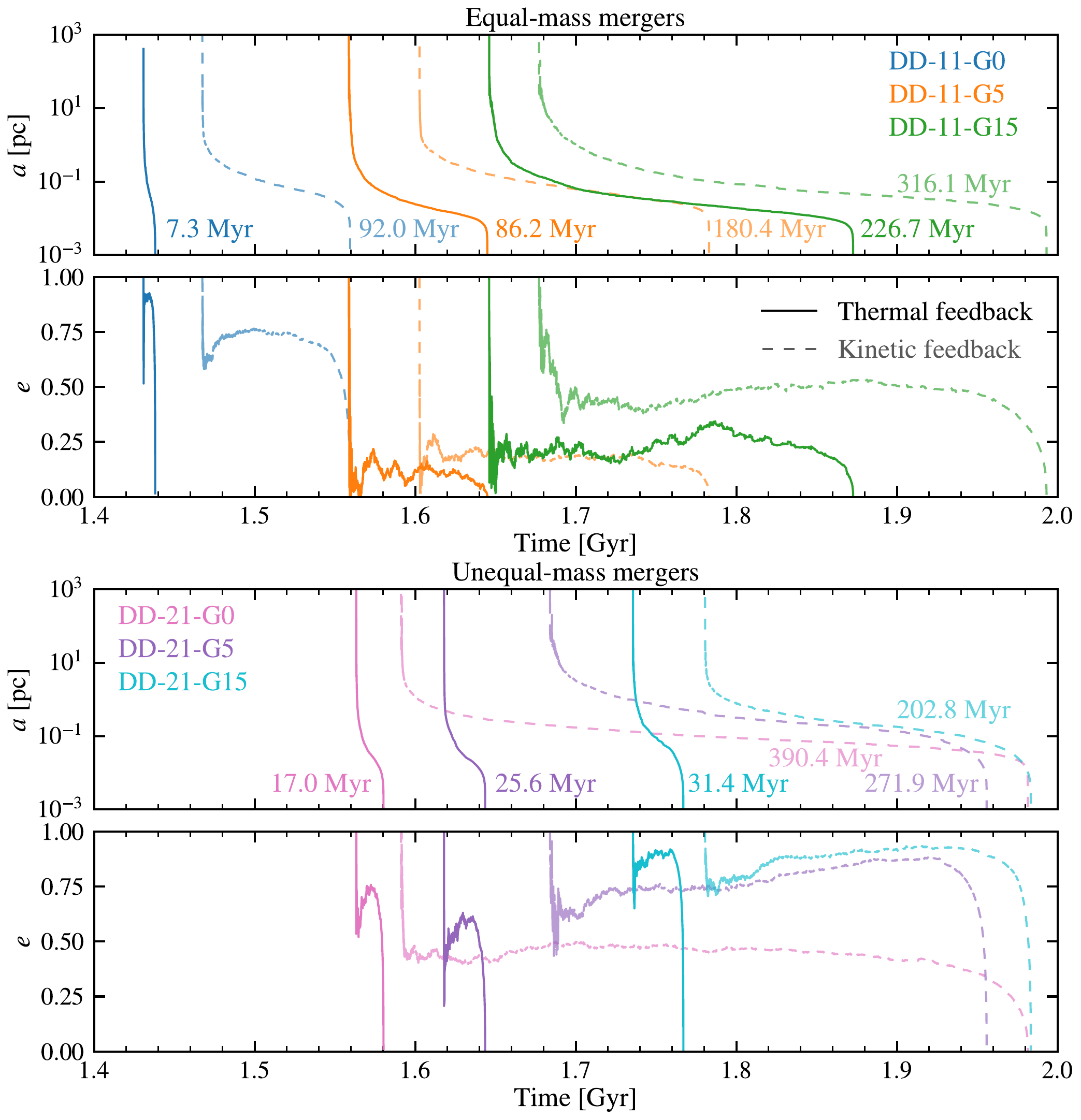}
\caption{Evolution of the PN corrected semi-major axis ($a$) and eccentricity ($e$) for the BH binaries in equal-mass (top) and the unequal-mass (bottom) galaxy mergers. Different galaxy mergers are shown with different colours as marked in the semi-major axis panels. The thermal and kinetic feedback runs are plotted using solid lines with dark colours and dashed lines with light colours, respectively. 
The corresponding BH merger time-scales are given in the semi-major axis plots.}
\label{fig:DD_orbit_params}
\end{figure*}

In Fig.~\ref{fig:DD_orbit_params}, we plot the PN corrected orbital parameters \citep{Memmesheimer2004,Mannerkoski2019}, the semi-major axis $a$ and the eccentricity $e$, for the BH binaries in different galaxy mergers (distinguished by colours) with both thermal (solid) and kinetic (dashed) feedback models. We also give the BH binary merger time-scales, which are measured from when the two BHs become a stably bound binary to when the two BHs merge.

For both equal-mass and unequal-mass mergers, the BHs in the G0 run form a bound binary first, then the G5 run, and only later the G15 run, reflecting that the BHs in the G0 run experience the strongest dynamical friction as a result of the strongest star formation in the galaxy centre \citep[see e.g.][]{Johansson2009ApJ}, in the G5 and G15 runs the merger induced starbursts are systematically weaker. Also, compared to equal-mass mergers, the BHs in the unequal-mass mergers with the same galaxy orbital geometry experience weaker dynamical friction, in agreement with previous work \citep[e.g.][]{Capelo2015}, and thus tend to form a binary later. For the same galaxy merger, the BHs in the kinetic feedback runs form binaries later than their counterparts in the thermal feedback runs, which is a result of the lower star formation rates in the kinetic feedback runs, which results in weaker dynamical friction.

All BH binaries in our simulations merge within a few hundred Myr, and there is no final parsec problem \citep{Begelman1980,Milosavljevic2003}. The merger time-scales in our simulations are in the range ${\sim} 10 - 400$ Myr. The DD-11-G0 run with thermal feedback has a particularly short merger time-scale of only ${\sim} 7~{\rm Myr}$. This equal-mass, co-planar, and prograde galaxy merger has the strongest tidal torque which drives a very high fraction of the gas to the galaxy centre resulting in a starburst, consequently, the interaction of these new stars and the BH binary leads to very efficient binary hardening. Note that in contrast to the gas-free galaxy merger simulations in which the merger time-scales of BH binaries with low eccentricities usually reach several Gyr \citep[e.g.][]{Khan2012a,Rantala2017}, the low-eccentricity BH binaries in our gas-rich galaxy mergers merge rapidly. For example, both the thermal and kinetic feedback runs of DD-11-G5 and the thermal feedback run of DD-11-G15 have relatively low eccentricities of ${\sim} 0.1$--$0.3$, but their merger time-scales are just ${\sim} 100$ to ${\sim} 200$ Myr. We also notice that overall the BH binary merger time-scales in the kinetic feedback runs (i.e. from ${\sim} 100$ to ${\sim} 400$ Myr) are longer than those in thermal feedback runs (i.e. from ${\sim} 10$ to ${\sim} 200$ Myr).

All the aforementioned results point to an explanation that in gas-rich disc galaxy mergers, the gas is driven to the galaxy centre and triggers the formation of new stars which replenish the loss cone of the BH binary; these stars interact with the BHs, and lead to an overall shorter merger time-scales compared to gas-free galaxy mergers. Compared to the thermal feedback case, the kinetic feedback is more effective in suppressing star formation during galaxy merging. Thus for comparable binary eccentricities, the BH binary merger time-scales in the kinetic feedback runs tend to be longer.

\subsection{Star ejections by the BH binary}

\begin{figure*}
\centering\includegraphics[width=450pt]{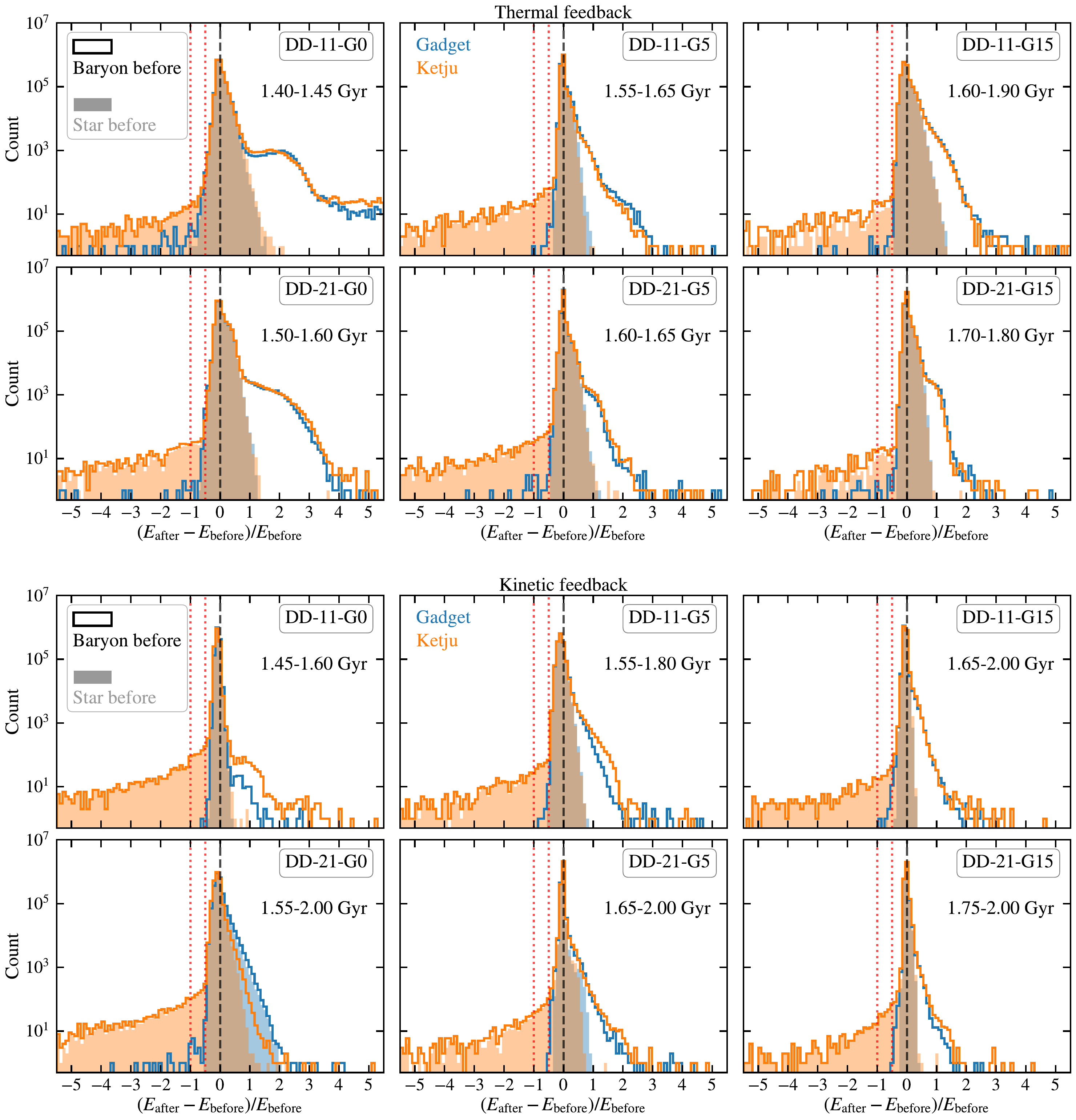}
\caption{Distribution of the specific energy difference $\Delta E / E \equiv (E_{\rm after} - E_{\rm before}) / E_{\rm before}$ in the thermal (top) and the kinetic (bottom) feedback runs. Different merger runs are plotted in different panels, as indicated by the legend at the top right corner. In each panel, the gadget and ketju + binary accretion runs are plotted with blue and orange colours, respectively. The open histograms plot the results of star particles in the snapshot after BHs merge (`after' snapshot) whose progenitors in the snapshot before BHs form a binary (`before' snapshot) are either gas or stars (i.e. baryons), while the filled histograms show the results of those star particles in the `after' snapshot whose progenitors in the `before' snapshot are stars only. The simulation time of the `before' and `after' snapshots are given in each panel using the format of `$t_{\rm before}$ -- $t_{\rm after}$ Gyr'. The vertical lines from left to right show $\Delta E / E = -1$, $-0.5$, and $0$.}
\label{fig:energy_diff}
\end{figure*}

\begin{table*}
\begin{threeparttable}
\caption{BH and galaxy properties from the snapshots just after BH mergers}
\label{tab:bh_info_after}
\begin{tabular}{lcccrcllccrc}
\hline
Name & AGN & $M_{\rm BH}$  & $M_{\rm ej}$ & $M_{\star}^{100~{\rm pc}}$ & $M_{\rm ej} / M_{\star}^{100~{\rm pc}}$ & $r_{\rm b}$ & $\gamma$ & $n$ & $r_{\rm e}$ & $\alpha^\prime$ & $\log_{10} \Sigma^\prime$\\
 & feedback & [$10^{8} {\rm M}_{\sun}$] & [$10^{8} {\rm M}_{\sun}$] & [$10^{8} {\rm M}_{\sun}$] & & [kpc] & & & [kpc] & & [${\rm M}_{\sun}{\rm kpc}^{-2}$]\\
\hline
DD-11-G0 & Thermal & 1.37 & 0.59 & 73.58 & 0.008 & 0.260 & 1.50 & 4.76 & 6.16 & 3.90 & 12.45 \\
DD-11-G5 & Thermal & 1.25 & 0.96 & 38.52 & 0.025 & 0.006 & 2.38 & 5.74 & 4.96 & 49.35 & 13.45 \\
DD-11-G15 & Thermal & 1.01 & 0.60 & 17.66 & 0.034 & 0.002 & 0.64 & 3.74 & 3.69 & 32.93 & 12.05 \\
DD-21-G0 & Thermal & 1.63 & 1.03 & 79.75 & 0.013 & 0.729 & 1.29 & 3.90 & 6.73 & 91.37 & 11.81 \\
DD-21-G5 & Thermal & 1.79 & 1.07 & 54.78 & 0.019 & 0.847 & 1.18 & 3.62 & 6.28 & 1.55 & 11.64 \\
DD-21-G15 & Thermal & 1.13 & 0.35 & 29.59 & 0.012 & 0.584 & 0.97 & 3.29 & 5.38 & 16.22 & 11.51 \\
\\
DD-11-G0 & Kinetic & 3.13 & 1.51 & 15.30 & 0.099 & 0.054 & 0.28 & 5.27 & 5.85 & 3.12 & 12.91 \\
DD-11-G5 & Kinetic & 3.09 & 0.83 & 8.70 & 0.096 & 0.043 & 0.96 & 3.59 & 4.57 & 46.97 & 11.73 \\
DD-11-G15 & Kinetic & 1.88 & 0.43 & 4.89 & 0.088 & 0.049$^\dagger$ & 0.10$^\dagger$ & 3.03 & 4.48 & 9.28 & 11.29 \\
DD-21-G0 & Kinetic & 4.00 & 2.31 & 5.24 & 0.441 & 0.102$^\dagger$ & 0.16$^\dagger$ & 3.69 & 6.17 & 3.90 & 11.72 \\
DD-21-G5 & Kinetic & 2.55 & 0.76 & 5.61 & 0.135 & 0.345 & 0.51 & 3.45 & 5.83 & 90.78 & 11.58 \\
DD-21-G15 & Kinetic & 1.61 & 0.38 & 5.02 & 0.077 & 0.052$^\dagger$ & 0.18$^\dagger$ & 2.97 & 5.55 & 157.11 & 11.22 \\
\hline
\end{tabular}
\begin{tablenotes}[flushleft]
 \footnotesize
 \item {\it Note}. The presented results are derived from the ketju + binary accretion runs. From left to right, the simulation name, the AGN feedback model used in the simulation, the BH mass, the ejected stellar mass, the enclosed stellar mass within 100 pc, the ratio between $M_{\rm ej}$ and $M_{\star}^{100~{\rm pc}}$, the break radius $r_{\rm b}$, the inner slope parameter $\gamma$, the outer shape parameter $n$, the effective radius $r_{\rm e}$, the transition sharpness parameter $\alpha^\prime$, and the normalisation $\Sigma^\prime$ from the core-S{\'e}rsic profile fitting. 
 \item $^\dagger$ The core radii and inner logarithmic slopes of core-like galaxy remnants.
\end{tablenotes}
\end{threeparttable}
\end{table*}

With the ketju integrator, we are able to resolve accurately the close dynamical interactions between star particles and the BH binary which cannot be resolved in traditional galaxy formation simulations (e.g. the gadget runs). Note that whilst gravitational slingshot ejections have been studied extensively in gas-free mergers \citep[e.g.][]{Milosavljevic2001,Merritt2006}, here by including gas physics and other galaxy formation processes, we will be able to address further related questions, such as: Are the ejected stars mostly young or old? Does the ejected stellar mass still correlate with the BH mass in hydrodynamic simulations, and do AGN feedback models affect the stellar ejections? In this subsection we try to answer these questions with our gas-rich galaxy merger simulations.

To study the slingshot kicks on star particles, we follow \citet{Frigo2021} and compute the change of the specific energy of star particles between the snapshot before the formation of a BH binary (the `before' snapshot) and the snapshot after the merger of BHs (the `after' snapshot), $\Delta E/E \equiv (E_{\rm after} - E_{\rm before}) / E_{\rm before}$, with $E_{\rm before}$ and $E_{\rm after}$ being the specific energy of a particle in the `before' and the `after' snapshots, respectively. Compared to the collisionless galaxy merger simulations used in \citet{Frigo2021}, our simulations contain gas and the related baryonic processes, and therefore additional effects need to be considered carefully when computing the energy change. Specifically: (i) a gas particle can be converted into a star particle in our simulations, and therefore the progenitor in the `before' snapshot of a star particle in the `after' snapshot can be either a gas or a star particle; (ii) as the over-massive gas particles are split into two particles to keep the mass resolution more uniform, we have to record the splitting history in our simulations so that we can trace the progenitors of any given baryonic (gas or star) particle in previous snapshots; (iii) as the masses of baryonic particles can change over time due to stellar feedback and metal diffusion, instead of computing the change of energy, here we compute the change of specific energy. The specific energy of a baryonic particle is computed as $E_i = \tilde{E}_i / m_{{\rm p}, i}$ where $\tilde{E}_i$ is the total energy and $m_{{\rm p}, i}$ is the particle mass. For a gas particle, $\tilde{E}_i$ is the sum of kinetic, potential, and internal energy, while for a star particle, $\tilde{E}_i$ is the sum of kinetic and potential energies.

The distributions of $\Delta E / E$ for different galaxy merger runs are shown in Fig.~\ref{fig:energy_diff}. To obtain this figure, we consider for each run all the star particles in the `after' snapshot, compute their $E_{\rm after}$, and trace back their progenitor particles (which can be either gas or star particles) in the `before' snapshot. For all bound\footnote{The fraction of unbound progenitor particles (i.e. gas particles which have recently received energy from feedback processes) within twice the stellar half-mass radius is negligible $(\la 10^{-5})$ for all runs.} progenitor particles which are within twice the stellar half-mass radius (centred on the CoM position of the BH binary\footnote{The centre of a merging galaxy is difficult to define while it is still in the merging process. Here we have tried three definitions for the galaxy centre: the CoM of the BH binary, the position of either BH, and the stellar centre computed with the shrinking sphere method \citep{Power2003}. The results are similar for all of the methods.}), we compute their $E_{\rm before}$, then compute the relative energy change $\Delta E / E$, and plot the distribution. Note that as $E_{\rm before}$ is negative, $\Delta E / E > 0$ means that the particle has become more bound while $\Delta E / E < 0$ means that it has become less bound or even unbound. To study the impact of the gas progenitor particles, we plot separately the histograms for all baryonic progenitor particles (open) and for only star progenitor particles (filled). 

For all runs, the histograms at $\Delta E/ E > 0$ are relatively similar for the gadget and ketju runs. The particles with very positive energy change (e.g. $\Delta E / E \ga 1$) are mostly gas progenitor particles which have cooled down, formed stars, and become more bound. In contrast, in the negative energy change range, the ketju runs have significant tails of particles extending to very negative values which have interacted with the BH binary and become less bound or even unbound. On the other hand, the gadget runs only have a small number of particles with $\Delta E / E < -1$, and they are usually gas progenitor particles which have received energy from feedback processes in the later evolution.

In the negative energy change range, the ketju histograms start to significantly differ from the gadget ones at $\Delta E / E \la -0.5$. Thus, we use $\Delta E / E < -0.5$ to approximately define the particles which have received strong kicks from the BH binary. These particles are pushed to larger radii (for $-1 \la \Delta E/E < -0.5$) or even escape from the galaxy (for $\Delta E / E < -1$), and thus they can affect the central dynamics of the galaxy. Note that as we can see from the gadget histograms, the baryonic feedback processes can also lead to negative energy changes for some particles, but as the number of such particles is very low compared to the star particles kicked by the BH binary in the ketju histograms, they do not affect our estimation in any significant way.

\begin{figure} 
\centering\includegraphics[width=\columnwidth]{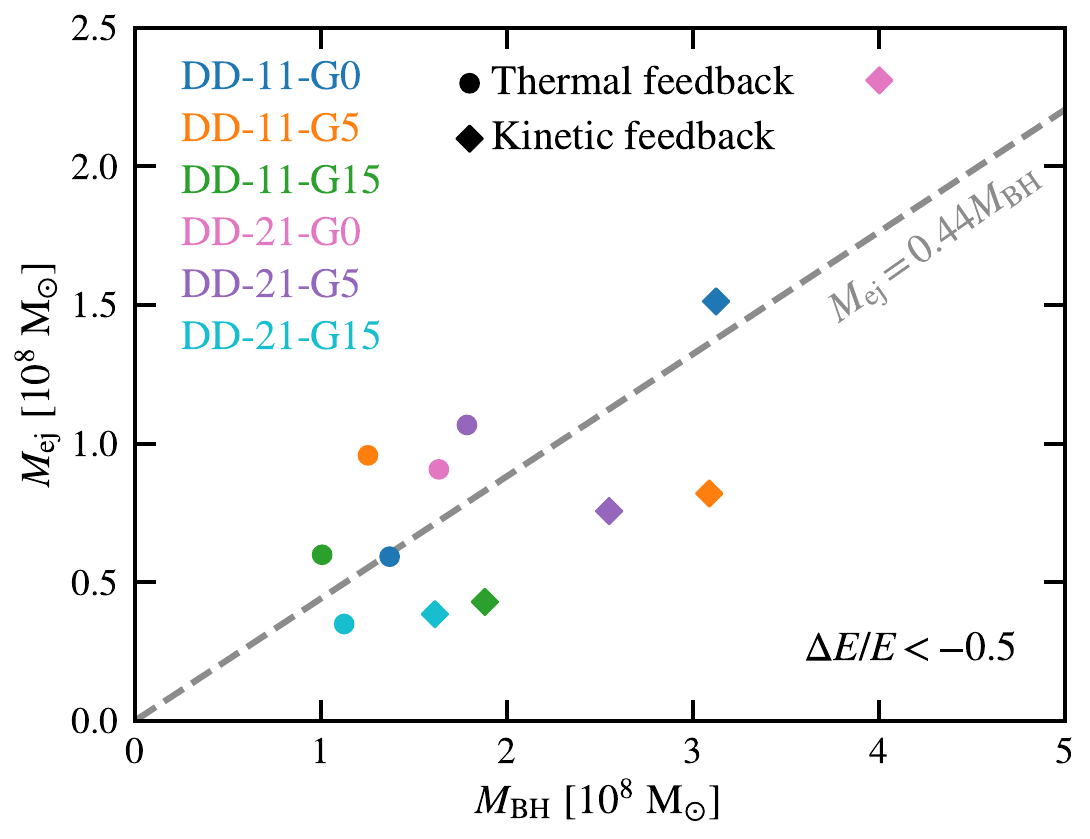}
\caption{The relation between the ejected stellar mass and the mass of the merged BH. Different galaxy mergers are distinguished by different colours, as shown in the legend. The thermal and kinetic feedback runs are plotted with filled circles and diamonds, respectively. The dashed line shows the best linear fit between the ejected stellar mass and the merged BH mass considering all data points from both the thermal and kinetic feedback runs, $M_{\rm ej} = 0.44 M_{\rm BH}$.}
\label{fig:ejected_mass}
\end{figure}

The total mass of the star particles with $\Delta E / E < -0.5$, i.e. the mass of ejected stars $(M_{\rm ej})$, and the mass of the BH after merging $(M_{\rm BH})$ are summarised in Table~\ref{tab:bh_info_after}, and their relation is shown in Fig.~\ref{fig:ejected_mass}. We can see that for both thermal and kinetic feedback runs, there is a positive correlation between $M_{\rm ej}$ and $M_{\rm BH}$, and a linear fit to all data points in Fig.~\ref{fig:ejected_mass} gives $M_{\rm ej} = 0.44 M_{\rm BH}$, i.e. the mass of the strongly ejected stars is roughly half of the total BH mass.

We notice that in Fig.~\ref{fig:ejected_mass}, the BHs in the kinetic feedback runs tend to be more massive than the BHs in the thermal feedback runs, which is due to two factors. First, as we have discussed in Fig.~\ref{fig:M_sigma_relation}, the final BH masses in our calibrated kinetic feedback model tend to be slightly more massive than those in the thermal feedback model. Second, note that the BH masses in Fig.~\ref{fig:ejected_mass} are measured from the `after' snapshots. Compared to the thermal feedback runs, the BH binaries in the kinetic feedback runs tend to have longer merger time-scales, and as a result, they have already accreted more mass and have merged to become more massive BHs in the `after' snapshots. Note that we also fitted the linear relation between $M_{\rm ej}$ and $M_{\rm BH}$ separately for the thermal and kinetic feedback runs, and the results, $M_{\rm ej} = 0.55 M_{\rm BH}$ and $M_{\rm ej} = 0.41 M_{\rm BH}$ respectively, are similar to the aforementioned combined fit, suggesting that the combined fit is a good representation of the overall correlation.

As a comparison, previous studies have investigated the stellar mass deficit in bright elliptical galaxies, $M_{\rm def}$, which is defined as the difference in the integrated mass between the assumed initial cuspy profile and the observed/simulated final cored profile. From observational data and gas-free galaxy merger simulations, it has been found that $M_{\rm def}$ correlates with the BH mass as $M_{\rm def} = f M_{\rm BH}$ with the factor $f$ varying from ${\sim} 0.5$ to ${\sim} 10$, and there have been suggestions that $f$ is proportional to the number of mergers \citep[e.g.][]{Milosavljevic2001,Milosavljevic2002,Graham2004,Merritt2006,Rantala2019,Nasim2021}. Note that we adopt here a definition that is different from these studies to quantify the mass of the stars which have strongly interacted with a BH binary, but interestingly we see a correlation between $M_{\rm ej}$ and $M_{\rm BH}$ similar to the one between $M_{\rm def}$ and $M_{\rm BH}$ in the single merger case \citep[i.e. $f=0.5$,][]{Merritt2006}. This similarity suggests that our adopted criterion (i.e. $\Delta E/E < -0.5$) for the definition of ejected stellar mass is reasonable.

\begin{figure} 
\centering\includegraphics[width=\columnwidth]{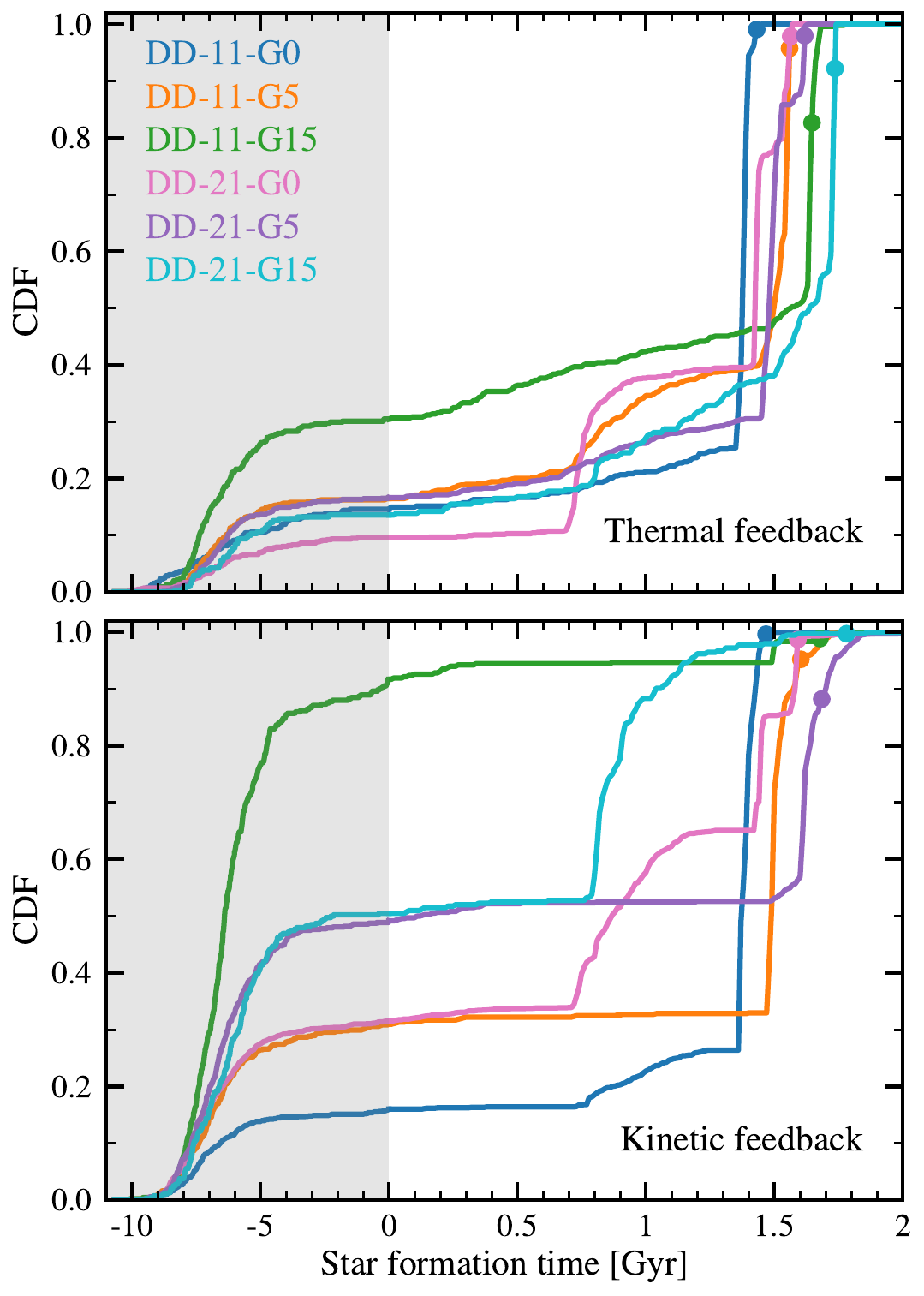}
\caption{Cumulative distribution function (CDF) of the star formation time of ejected star particles for the thermal (top) and kinetic (bottom) feedback runs. Note that the scales of the negative (marked by the grey background) and the positive $x$-axis are different. Star particles with negative star formation time are already present in the initial condition and their formation time are generated using the methods described in Section~\ref{subsec:sim_ic}. Lines with different colours show the results of different galaxy mergers. The filled circle associated with each line marks the onset of the BH binary phase in the corresponding simulation.}
\label{fig:ejected_ages}
\end{figure}

In Fig.~\ref{fig:ejected_ages}, we show the cumulative distribution functions (CDFs) of the formation time of the strongly ejected stars in the different runs. Star particles with negative formation times are already present in the initial condition and their formation times are generated using the methods described in Section~\ref{subsec:sim_ic}. For all thermal feedback runs, there are sharp increases of CDFs at roughly $200~{\rm Myr}$ prior to the onset of the BH binary phase, indicating that the majority of the strongly ejected stars are newly formed stars. These new stars form from the gas driven to the galaxy centre by the tidal torques during the galaxy mergers. In the DD-11-G0 run which has the strongest starburst, nearly $80$ per cent of the ejected stars are formed after $t = 1.2~{\rm Gyr}$ (i.e. ${\sim} 200~{\rm Myr}$ prior to the formation of the bound BH binary). For the other orbital geometries, more than $50$ per cent of the ejected stars are formed within the $200~{\rm Myr}$ time window of the onset of the BH binary phase. Qualitatively similar behaviour is also seen in the kinetic feedback runs. However, as kinetic AGN feedback is more effective in suppressing the star formation compared to thermal feedback (see Appendix~\ref{ap:kin_fb}), the fractions of newly formed stars in the ejected stellar population are lower.

\subsection{Stellar density profiles}\label{subsec:density_profile}

\begin{figure*} 
\centering\includegraphics[width=450pt]{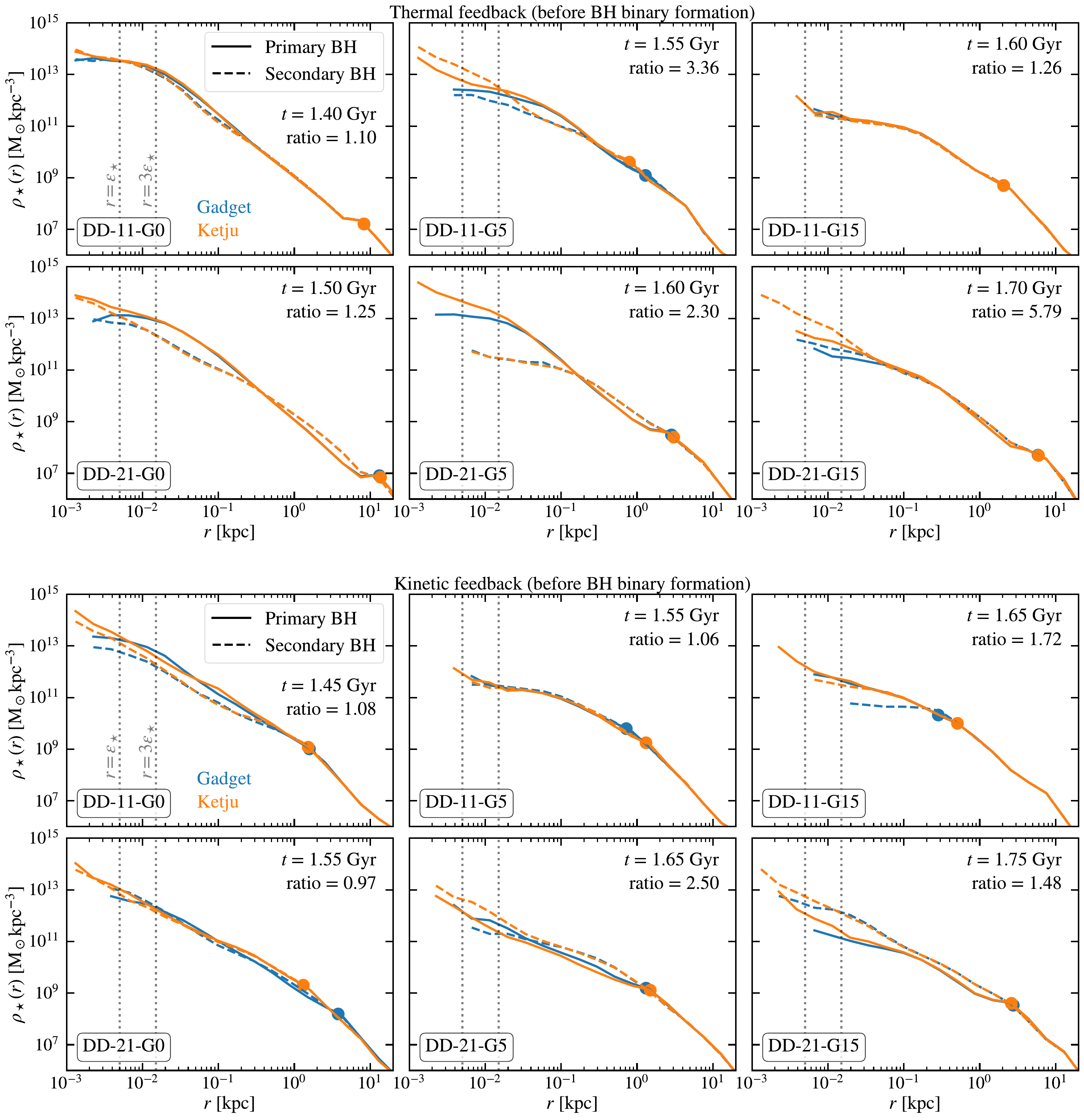}
\caption{Stellar density profiles from the snapshots before the BHs form a binary in the thermal (top) and kinetic (bottom) feedback runs. Different panels show the results from different galaxy mergers, as indicated by the legend at the bottom left corner of each panel. In each panel, the gadget and ketju + binary accretion runs are shown with blue and orange colours respectively. The solid and dashed lines plot the stellar density centred on the primary and secondary BHs, respectively. The filled circle marks the distance between the primary and secondary BHs. Note that we only plot the density profile bins with at least $5$ particles, and thus the innermost radial bins vary for different profiles. The vertical dotted lines mark the softening length of the star particles $\epsilon_{\star}$ and the ketju region radius $r_{\rm ketju} = 3\epsilon_{\star}$. At the top right corner of each panel, the simulation time of the `before' snapshot and the total enclosed mass ratio between the ketju and the gadget runs within the ketju radius are provided.}
\label{fig:density_profile_before}
\end{figure*}

\begin{figure*} 
\centering\includegraphics[width=450pt]{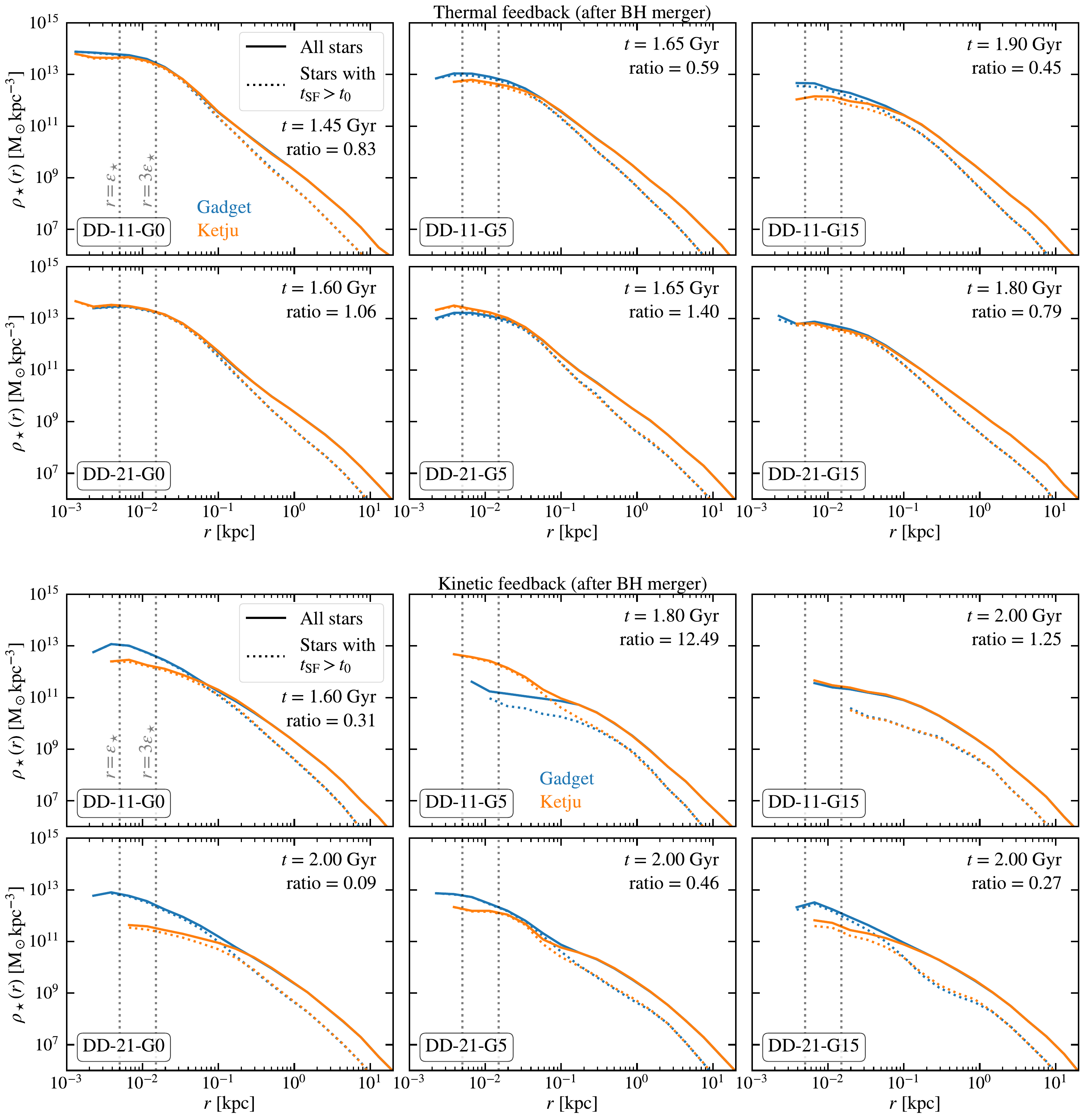}
\caption{Similar to Fig.~\ref{fig:density_profile_before}, but for the stellar density profiles from the snapshots after the BH merger. The profiles are centred on the merged BH. The solid and dotted lines plot the stellar density computed from all the star particles and the stars formed after the start of the simulation (i.e. with star formation time $t_{\rm SF} > t_0$), respectively. The simulation time of the `after' snapshot and the enclosed mass ratio within the ketju radius between the ketju and the gadget runs are given at the top right corner of each panel.}
\label{fig:density_profile_after}
\end{figure*}

In the ketju runs the BH binary can eject star particles via the slingshot interaction with the total amount of ejected mass roughly equalling half of the total BH mass. Thus, it is interesting to compare the galaxy stellar density profiles in the ketju and gadget runs and study how the star ejection affects the central stellar distribution. It will also be informative to compare the profiles with the simulations of gas-free massive elliptical galaxy mergers, which have shown that the ejection of stars by the BH binary can effectively lower the central stellar density profiles and produce cores with sizes ranging from a few hundred pc up to a few kpc \citep[e.g.][]{Rantala2018,Nasim2021}.

In Fig.~\ref{fig:density_profile_before}, we plot the stellar density profiles centred on the primary (solid lines) and secondary (dashed lines) BHs in the `before' snapshots from both the thermal (upper panels) and kinetic (lower panels) feedback runs. Overall, before the BHs form a bound binary, the stellar density profiles at radii beyond a few stellar softening lengths are quite similar between the gadget (blue colour) and ketju (orange colour) runs, especially for the thermal feedback runs. In the kinetic feedback runs, sometimes the ketju and gadget stellar density profiles can show observable differences at larger radii (e.g. the profiles centred on the primary BHs in the DD-11-G0, DD-21-G0, and DD-21-G5 runs and the one centred on the secondary BH in run DD-11-G15), which should originate from the higher stochasticity of the pulsed kinetic feedback compared to the more continuous thermal feedback. At radii of $r \la r_{\rm ketju} = 3 \epsilon_{\star}$, in both the thermal and kinetic feedback simulations, the stellar density profiles in the ketju runs are cuspier than in the gadget runs. As a quantitative measurement, we compute the summed enclosed stellar mass within the ketju region radius $r_{\rm ketju}$ centred on both BHs and show the mass ratio between the ketju and gadget runs in the top-right corner of each panel in Fig.~\ref{fig:density_profile_before}. We can see that this ratio tends to be above one, and in some runs it can even reach a factor of a few. This difference in the very inner density profile can be explained by the fact that in the ketju runs, non-softened gravity is considered between BHs and stars within the regularised region while in the gadget runs, softened gravity is used within the distance of $2.8\epsilon_{\star}$ for all interactions, leading to artificially more flattened profiles. 

The stellar density profiles centred on the merged BH in the `after' snapshots are shown in Fig.~\ref{fig:density_profile_after}. The solid lines show the density profiles computed with all star particles while the dotted lines give the density profiles for stars formed after the start of the simulation (i.e. with star formation time $t_{\rm SF} > t_0$). Despite the fact that in the `before' snapshot the stellar densities around ketju BHs tend to be slightly higher than around the gadget BHs in the very central region, after the two BHs merge, most of the ketju galaxy remnants tend to have similar or even lower stellar densities in the central region compared to their gadget counterparts. This reflects the effects of the ejection of stars by the BH binary via the slingshot interaction. This effect is more significant in the kinetic feedback runs because their ejected stellar masses are higher (see Fig.~\ref{fig:ejected_mass}) and their central stellar densities are lower compared to their counterparts in the thermal feedback runs. At the top-right corner of each panel, we again give the ketju-to-gadget ratio of the enclosed stellar mass within $3\epsilon_{\star}$. Note that the ratios of some merger remnants are greater than 1 (e.g. thermal DD-21-G0 and DD-21-G5, kinetic DD-11-G15), however, compared to the corresponding ratios in the `before' snapshots (Fig.~\ref{fig:density_profile_before}), they have decreased, suggesting that the star ejection in the ketju run is indeed lowering the central stellar densities. 

However, we can clearly notice one outlier, the DD-11-G5 kinetic feedback simulation, which seems to contradict the conclusions above. For this simulation, after the two BHs merge, the central stellar density profile in the ketju run is significantly higher than that in the gadget run, even though the stellar density profiles in these two runs are very similar in the `before' snapshots. After an examination of the star formation histories in these two runs, we find that the central density bump in the ketju run originates from the formation of an ultra-compact stellar clump, with a stellar mass of ${\sim} 5 \times 10^{8}~{\rm M}_{\sun}$ and a stellar half-mass radius of ${\sim} 20$ pc. At $t \sim 1.5$ Gyr, the small differences in the gas distribution in the tidal arm in the ketju and gadget runs lead to small differences in the star formation which is a stochastic process. In the ketju run, the initially formed stellar clumps are more compact and clustered, and in the subsequent evolution, the small differences are amplified in a runaway manner as the clumps merge to form an ultra-compact clump which is compact enough that it survives the tidal interactions and finally merges into the centre of the galaxy merger remnant, thus boosting significantly the central density profile and overshadowing the effect of star ejection. 

In contrast, in the gadget run these initially formed stellar clumps in the tidal arm are less compact and less clustered. In their later evolution they remain as small and separate clumps, and are finally tidally disrupted when they merge into the galaxy remnant centre. We have performed a test gadget run starting with the ketju `before' snapshot at $t = 1.55$ Gyr which contains the ultra-compact clump moving along the tidal arm, and found that at $t = 1.80$ Gyr, the stellar density profile in this test gadget run is indeed higher than the ketju one, which is similar to the other galaxy mergers discussed before. See Appendix~\ref{ap:dd_11_g5} for more details.

In order to quantify if a ketju galaxy merger remnant has a cored density profile, we compute the surface density profile $\Sigma (r)$ in the `after' snapshot for each ketju + binary accretion run. We project the star particles along 50 random viewing directions, compute $\Sigma (r)$ in the projected radius range of $[0.01, 60]$ kpc for each viewing direction, stack $\Sigma (r)$ over all viewing directions, and fit the stacked profile using the six-parameter core-S{\'e}rsic profile \citep{Graham2003,Trujillo2004}
\begin{equation}
    \Sigma (r) = \Sigma^\prime \left[1 + \left(\frac{r_{\rm b}}{r}\right)^{\alpha^\prime}\right]^{\gamma/\alpha^\prime} \exp\left[-b \left(\frac{r^{\alpha^\prime} + r_{\rm b}^{\alpha^\prime}}{r_{\rm e}^{\alpha^\prime}}\right)^{1/(\alpha^\prime n)}\right],
\end{equation}
where $\Sigma^\prime$ is the normalisation surface density, $r_{\rm b}$ is the break radius separating the inner power-law with a logarithmic slope $\gamma$ and the outer S{\'e}rsic function with an index $n$ and an effective radius $r_{\rm e}$, $\alpha^\prime$ is a parameter controlling the sharpness of the break, and $b$ is defined by requiring that $r_{\rm e}$ becomes the radius enclosing half of the total mass of the profile. The best-fitting parameters for the different runs are summarised in Table~\ref{tab:bh_info_after}. Note that the fitted profiles with very small $r_{\rm b}$ (i.e. less than the lower limit of the fitting radius range, $0.01$ kpc) are essentially S{\'e}rsic profiles \citep{Sersic1963}, see e.g. the thermal runs of DD-11-G5 and DD-11-G15.

In the thermal feedback runs, the inner logarithmic slopes are $\gamma \ga 1$, suggesting that all galaxy remnants are cuspy galaxies. In the kinetic feedback runs, galaxy remnants tend to have smaller $\gamma$. Especially, the DD-11-G15, DD-21-G0, and DD-21-G15 remnants, for which the break radii and inner slopes are marked with dagger symbols ($\dagger$) in Table~\ref{tab:bh_info_after}, have $\gamma < 0.2$, suggesting that these remnants have core-like stellar density distributions with core radii (i.e. the break radii) ranging from ${\sim} 50$ pc to ${\sim} 100$ pc. In Table~\ref{tab:bh_info_after}, we also provide the central enclosed stellar masses within $100$ pc in the galaxy remnants, $M_{\star}^{\rm 100~pc}$, and the ratios between the ejected stellar mass and this central stellar mass. In the thermal feedback runs, the ejected stellar masses are usually only $\la 3$ per cent of the central stellar masses while in kinetic feedback runs, this ratio is larger, reaching $\ga 10$ per cent. Therefore, it is not surprising that in the thermal feedback runs we do not see core formation as the ejected stellar mass is almost negligible compared to the total stellar mass in the central region. Our results indicate that the AGN feedback model can play an important role in the formation of cores in gas-rich galaxy mergers. 

Compared to the gas-free galaxy merger simulations, which produce large cores with radii of a few hundred parsecs to even a few kiloparsecs \citep[e.g.][]{Rantala2018,Nasim2021}, our galaxy remnants are mostly cuspy galaxies or cored galaxies with small cores (i.e. $r_{\rm b} \la 100$ pc). The following arguments explain this difference. Firstly, in gas-free simulations of massive elliptical galaxies, the BH masses are usually much higher (e.g. $M_{\rm BH} \sim 10^{9}$--$10^{10}~{\rm M}_{\sun}$) which will lead to more significant stellar ejections and thus assist in the formation of larger cores. Secondly, unlike gas-free simulations, there are new stars forming in the galaxy central region, which will dominate the central density in gas-rich galaxy mergers (see the dashed lines in Fig.~\ref{fig:density_profile_after}). This replenishment of stars can hinder the formation of a large core, i.e. these new stars will interact with the BH binary and drive it to merge in a shorter time-scale, meanwhile they can also increase the central stellar density, both of which can make the core formation mechanism less effective. Especially in the thermal feedback runs, star formation is stronger and the central stellar profiles remain cuspy.

Finally, we note that observations have revealed that bright elliptical galaxies ($M_V \la -22$) tend to have core-like surface brightness profiles, while fainter elliptical galaxies ($M_V \ga -20$) tend to have steeper profiles that lack cores, and at intermediate magnitudes, both types coexist \citep[e.g.][]{Faber1997,Trujillo2004,Lauer2007}. As our simulated galaxy remnants are elliptical galaxies roughly in the intermediate-mass range, the results that only some of them exhibit small cores and some of them even have no cores are consistent with these observations. For our simulated galaxy remnants which exhibit cores, they broadly agree with the observed relation between BH masses and core radii provided in \citet{Thomas2016}.

\section{Discussions} \label{sec:dis}

In this section, we discuss the caveats of our BH binary accretion and feedback model and the improvements that we can introduce in the future.

{\it (i) Fitting formulae from circumbinary disc simulations.} In our binary BH model, the preferential mass accretion behaviour is included by adopting the fitting formula from circumbinary disc simulations. However, all the formulae shown in Fig.~\ref{fig:cbd_fit_eq} are fitted from simulations with circular ($e=0$) BH binaries, and therefore strictly speaking, it is not fully self-consistent to apply these formulae to eccentric BH binaries in our simulations. There have been circumbinary disc simulations of eccentric BH binaries \citep[e.g.][]{Artymowicz1996,Hayasaki2007,Cuadra2009,Roedig2011,Munoz2019,Zrake2021}, but there are still no fitting formulae available for such eccentric cases and the sampled parameters, such as the mass ratio and eccentricity, are still fairly limited in these studies \citep[but see the recent work of][for a systematic exploration of the binary mass ratio and eccentricity parameter space]{Siwek2022}. Depending on the different binary mass ratios, the binary eccentricity might lead to suppressed or enhanced preferential accretion compared to the circular case \citep[see][]{Siwek2022}. 

Thus, how the BH accretion rate ratio depends on both the binary mass ratio and the eccentricity is still an ongoing research topic and hopefully a robust fitting formula will be provided from the community in the near future. The fitting formula implemented in our code is a module which can be easily changed and updated, and it will be straightforward to incorporate an improved formula. We note that similar to our approach, the fitting formulae from circular BH binaries have also been adopted in some semi-analytical modelling studies, see e.g. \citet{Kelley2019}.

We further caution that the existing fitting formulae are usually based on two-dimensional simulations which assume that the BH binary is coplanar with the circumbinary disc. The simulations also assume a specific disc thickness and viscosity, and also ignore the effect of AGN feedback. Note that much of the parameter space of circumbinary disc simulations remains unexplored \citep{Lai2022}. See recent works of \citet{Dittmann2022} for a systematic exploration of the disc thickness and viscosity, and finally see \citet{delValle2018} for a study on the effects of AGN feedback on the evolution of the BH binary + circumbinary disc system.

{\it (ii) Circumbinary disc-BH binary torque interactions.} In a circumbinary disc-BH binary system, apart from the mass accretion, the gravitational torque from the disc could also affect the orbital evolution of the BH. The torque experienced by the binary (i.e. the rate of change of the angular momentum of the binary due to disc-binary interaction, $\dot{J}_{\rm bin}$) is composed of two parts, i.e. $\dot{J}_{\rm bin} = T_{\rm grav} + \dot{J}_{\rm acc}$, where $T_{\rm grav}$ is the gravitational torque from all fluid elements (including the circumbinary disc, streams, and mini-discs), and $\dot{J}_{\rm acc}$ is the angular momentum change due to accretion. Note that even if the accreted gas has zero total angular momentum, adding its mass to the binary can increase the angular momentum of the binary, and therefore a part of $\dot{J}_{\rm acc}$ has been modelled automatically in our subgrid model. 

However, the modelling of $T_{\rm grav}$ is a much more complicated story. Early analytical and numerical works suggest that $T_{\rm grav}$ is negative, i.e. the BH binary loses angular momentum to the disc and shrinks its orbit, thus the disc-binary interaction assists the BH coalescence \citep[e.g.][]{Artymowicz1991,Pringle1991,MacFadyen2008}. In contrast, several recent numerical works investigating the disc-binary interaction in more detail show that $T_{\rm grav}$ is positive \citep[][]{Miranda2017,Tang2017,Moody2019,Munoz2019,Duffell2020,Munoz2020,Dittmann2021}, i.e. although the gravitational torque from the circumbinary disc (region with radius $r \ga 2a$ with $a$ being the binary semi-major axis) is negative (as expected in early works), the torque from the gas in the innermost region ($r \la a$) is overwhelmingly positive, leading to a positive total torque $T_{\rm grav}$. This implies that the binary gains net angular momentum and its orbit expands, suggesting that the circumbinary disc may prevent rather than assist the BH merger. Follow-up simulations further find that the positive $T_{\rm grav}$ from these numerical works might originate from their assumption of a thick circumbinary disc, i.e. a disc with an aspect ratio (i.e. the ratio between the disc thickness $h$ and radius $r$) of $h/r = 0.1$ or equivalently Mach number of $\mathcal{M}=10$ \citep{Heath2020,Tiede2020,Dittmann2022}. These authors show that with a thinner disc (e.g. $h/r \la 0.04$ or $\mathcal{M} \ga 25$), which is more preferred for AGN, the inner edge of the circumbinary disc is denser and its negative torque can dominate; as a result, the total torque $T_{\rm grav}$ is negative and the binary migrates inwards instead of outwards. Note that $T_{\rm grav}$ might also depend on the viscosity \citep{Dittmann2022} and the sink prescription \citep{Dittmann2021}. Given all these complexities and the on-going debates in the literature, unlike the preferential mass accretion, the modelling of $T_{\rm grav}$ is much more uncertain and difficult, and we do not include it in our current subgrid accretion model. We note that even without including the full disc-binary interacting torque in our simulations, all simulated BH binaries merge within a few hundred Myr, suggesting that the star formation in gas-rich galaxy mergers is efficient enough in driving the binary to the regime of GW emission through slingshot interactions. It will be interesting to incorporate the disc-binary torque interaction into our subgrid model in the future when a sophisticated modelling of $T_{\rm grav}$ is available from circumbinary disc simulations.

{\it (iii) Accretion of gas angular momentum.} In this work we focus on the mass accretion of BHs and ignore the accretion of gas angular momentum and the resulting evolution of BH spins, which are important when modelling spin-driven jet AGN feedback and the GW recoil kicks at the last stage of BH coalescence. Recently, there have been a range of attempts to improve the BH subgrid model by taking into account the BH spins. For example, mass accretion and spin evolution based on the Shakura–Sunyaev $\alpha$-disc model has been considered in e.g. \citet{Dubois2014,Fiacconi2018,Bustamante2019,Cenci2021}, and the spin-driven jet AGN feedback have been implemented in e.g. \citet{Talbot2021,Husko2022}. Note that the circumbinary disc can also affect the spins of the merging BHs \citep[e.g.][]{Steinle2022}. We leave the construction of a subgrid model for BH spin evolution due to gas accretion for future work.

{\it (iv) Two-mode feedback model.} Usually, a combined two-mode AGN feedback (for example, thermal/kinetic feedback for quasar/radio mode) is preferred in galaxy formation simulations as this is motivated by observations and has been shown to be useful in reproducing the observed galaxy properties in modern cosmological simulations \citep[e.g.][]{Sijacki2007,Dubois2012,Hirschmann2014,Steinborn2015,Grand2017,Weinberger2017,Dave2019}. Here, as a first attempt to introduce a BH binary accretion model, we aim to study if and how different AGN feedback implementations affect the BH binary evolution, and thus we use either a pure thermal or a pure kinetic feedback model in our simulations. Similar approaches have been adopted in previous simulation studies of isolated or merging galaxies \citep[e.g.][]{Barai2014,Barai2016}. Note that a two-mode model will introduce more parameters (e.g. the accretion rate criterion to switch between the two modes) and usually, it is difficult to calibrate such additional parameters with idealised galaxy merger simulations. We leave the introduction of a two-mode AGN feedback model for ketju simulations for future work.

\section{Conclusions}\label{sec:con}

In this work, we introduce a BH binary accretion and feedback model for the \ketju code, which enables us to resolve the entire BH coalescence process in gas-rich galaxy merger simulations in a more physically motivated way.

Our BH binary accretion model is a natural extension of the BHL accretion model \citep{Hoyle1939,Bondi1944,Bondi1952}, which has been widely used in galactic-scale simulations \citep{DiMatteo2005,Springel2005BH}, and it also incorporates the results from small-scale viscous hydrodynamical simulations of circumbinary discs \citep{Farris2014,Duffell2020}. Specifically, our model adopts a simple switch between the single and binary BH phases. In the single BH phase, we follow the traditional BHL accretion model, whereas in the binary phase, a circumbinary disc subgrid model is introduced. When two BHs form a bound binary, the total accretion rate onto the circumbinary disc is computed with the BHL accretion model but using the gas properties at the binary centre-of-mass, while the fitted formula from small-scale circumbinary disc simulations is adopted to assign the accretion to each BH. The AGN feedback is implemented with both thermal and kinetic approaches. In the binary phase, the kick velocity direction of the kinetic feedback is set to be parallel or anti-parallel to the orbital angular momentum direction of the BH binary.

To test and demonstrate the behaviour of the model, we perform idealised galaxy merger simulations with two gas-rich disc galaxies and consider both pure thermal AGN feedback and pure kinetic AGN feedback in our runs. Our major findings are summarised as follows.

{\it (i) Accretion behaviour of BH binaries.} We demonstrate that using the centre-of-mass properties of the BH binary in the BHL formula is important when estimating the accretion onto the BH binary, otherwise the high inspiral BH velocities in the tight binary phase will artificially suppress the BH accretion rates and the AGN feedback strength, and lead to unphysical abrupt jumps in the evolution of accretion-related quantities. Furthermore, as a result of the preferential mass accretion from circumbinary disc simulations, the BH binaries tend towards equal-masses in our simulations. Note that such an improved modelling for BH binary accretion is of paramount importance, especially for gas-rich galaxy mergers which have higher accretion rates, as the BH binary accretion behaviour directly affects the BH mass ratio before coalescence which is one of the key parameters in computing the GW-induced recoil velocity \citep{Campanelli2007}.

{\it (ii) BH merger time-scales.} All BH binaries in our simulation merge rapidly, and the merger time-scales range from ${\sim} 10$ up to ${\sim} 400$ Myr. In gas-rich galaxy mergers, the tidal torque drives a large amount of gas to the galaxy centre, triggering starbursts which can replenish the loss cone of the BH binary and lead to overall shorter time-scales compared to gas-free galaxy mergers \citep[especially for binaries with lower eccentricities, e.g.][]{Khan2012,Rantala2018}. In our simulations, the orbital geometry with stronger starbursts (i.e. the co-planar prograde equal-mass merger G0) tends to have shorter BH merger time-scales (e.g. ${\sim} 10$ Myr in the thermal feedback model and ${\sim} 100$ Myr in the kinetic feedback model). Overall, in the kinetic AGN feedback runs, BHs merge over longer time-scales (i.e. from ${\sim} 100$ to ${\sim} 400$ Myr, compared with the time-scales from ${\sim} 10$ to ${\sim} 200$ Myr in the thermal feedback runs), as kinetic feedback quenches star formation more efficiently.

{\it (iii) Dynamical ejection of stars.} We use the change in the total energy of star particles before and after the binary phase to identify star particles which experienced strong dynamical kicks by the BH binary. We find that the total mass of the ejected star particles is roughly half of the total BH mass, agreeing with a similar correlation found from observations and gas-free galaxy merger simulations \citep{Milosavljevic2002,Merritt2006}. We also show that the ejected stars tend to have young stellar ages.

{\it (iv) Formation of cuspy and cored elliptical galaxies.} Due to the star ejection by the BH binary, the central stellar density profiles of our galaxy remnants tend to be lower compared to those in traditional gadget simulations which do not resolve the BH binary dynamics. However, unlike gas-free simulations of massive elliptical galaxy mergers \citep{Rantala2018,Nasim2021}, there are new stars forming in the galaxy central region, which increase the central density in gas-rich galaxies, and can thus hinder the formation of large cores. As a result, all merger remnants in the thermal feedback runs are cuspy galaxies because of the strong star formation activity. A few merged galaxies in the kinetic feedback runs exhibit core-like surface density profiles with small cores (i.e. core radii range from ${\sim}50$ to ${\sim}100$ parsecs). Overall our results agree with the observed elliptical galaxies at the intermediate magnitude range ($-22 \la M_V \la -20$) in which cuspy and cored ellipticals coexist \citep{Faber1997}.

{\it (v) Impacts of different AGN feedback implementations.} Both our pure thermal and pure kinetic feedback simulations reproduce the observed galaxy scaling relations such as the $M_{\rm BH}$--$\sigma_\star$ relation, the galaxy size--stellar mass relation, and the hot gas X-ray luminosity--stellar mass relation. As summarised above, we also notice that AGN feedback models play an essential role for setting the BH merger time-scales and for the formation of cores. This indicates that AGN feedback models are a crucial ingredient in modelling the BH binary coalescence in a cosmological galaxy formation context.

Our newly introduced BH binary accretion and feedback model in the \ketju code can be further applied to cosmological simulations by adding a BH seeding scheme \citep{Sijacki2007} as in \citet{Mannerkoski2022} and extending the model to multiple BH systems. BH mergers in late-type gas-rich galaxies are one of the major targets for future spaceborne GW observatories, such as LISA \citep{Amaro-Seoane2022} and TianQin \citep{Luo2016}. Our model will be useful for improving the modelling of these targets in galaxy formation simulations.

\section*{Acknowledgements}
We thank Thorsten Naab for valuable discussions and the anonymous referee for helpful comments. SL, PHJ, DI, and FPR acknowledge the support by the European Research Council via ERC Consolidator Grant KETJU (no. 818930). PHJ and TS acknowledge the support of the Academy of Finland grant 339127. TS also acknowledges the supports from Academy of Finland grants 311049 and 335607, and from the European Research Council (ERC) Advanced Investigator grant DMIDAS (GA 786910). The numerical simulations used computational resources provided by the CSC - IT Center for Science,
Finland.

We gratefully thank the developers of the open-source \textsc{Python} packages that were used in the data analysis of this work, including \textsc{Matplotlib} \citep{Hunter2007}, \textsc{NumPy} \citep{Harris2020}, \textsc{SciPy} \citep{Virtanen2020}, \textsc{Astropy} \citep{astropy2013,astropy2018,astropy2022}, and \textsc{Pygad} \citep{Rottgers2020}.

\section*{Data availability}
The simulation data used in this article will be shared upon a reasonable request to the corresponding author.

\bibliographystyle{mnras}
\bibliography{ref} 

\appendix

\section{Tests on the kinetic feedback model}\label{ap:kin_fb}

\begin{figure*}
\centering\includegraphics[width=425pt]{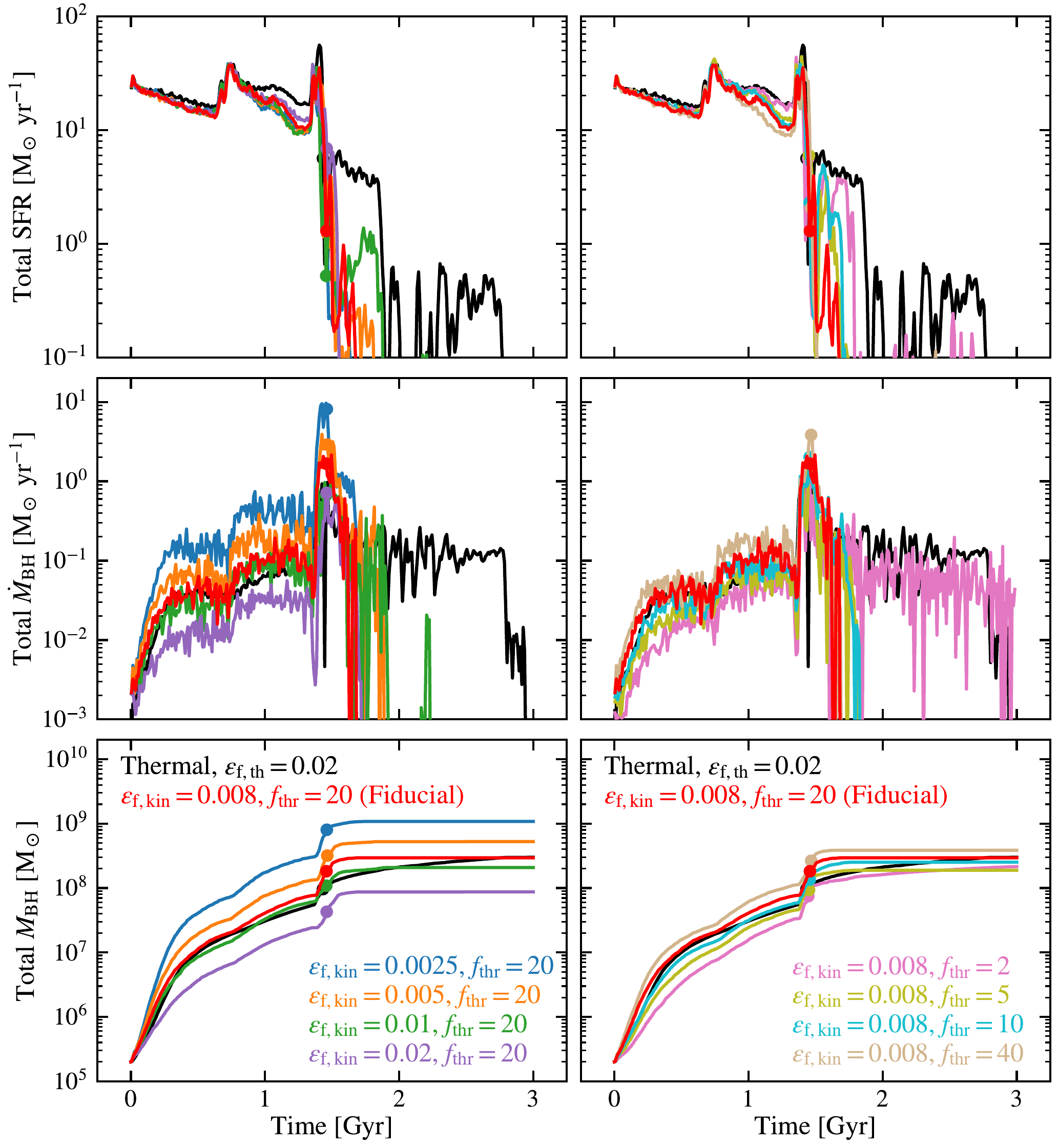}
\caption{The star formation and BH growth histories in the DD-11-G0 merger runs with different kinetic feedback model parameters. From top to bottom panels, the time-evolution of the total SFR, the total BH accretion rate, and the total BH mass are plotted. In the left column, the results from different kinetic feedback runs with varying $\epsilon_{\rm f, kin}$ (but with fixed $f_{\rm thr} = 20$) are shown, whereas in the right column, the results from different kinetic feedback runs with varying $f_{\rm thr}$ (but with fixed $\epsilon_{\rm f, kin} = 0.008$) are given. In all panels, the run with thermal feedback is plotted with black lines, while the run with our fiducial kinetic feedback model parameters (i.e. $\epsilon_{\rm f, kin} = 0.008$ and $f_{\rm thr} = 20$) is shown with red lines. The filled circles mark the merging of the BHs. Note that the total SFR and the total BH accretion rate are averaged over 10 Myr in this figure.}
\label{fig:ap_kin_fb_sfr_bhar}
\end{figure*}

In this appendix, we test the impact of different parameter values on our pure kinetic feedback model and find the parameter values which can reproduce the considered observations. To save computational time, we perform all test runs in this appendix using the gadget mode. As we have shown in Section \ref{sec:res}, when the parameter values calibrated with the gadget runs are applied to the ketju + binary accretion runs, the simulated scaling relations agree well with the observations. This is not surprising since the ketju runs have very similar overall BH growth and star formation histories as compared to the gadget runs (Section~\ref{subsec:bh_growth_his}).

All tests are run with the DD-11-G0 galaxy merger, and except for the BH model parameters, all other parameters are identical in all runs. To study the effects of the feedback efficiency parameter $\epsilon_{\rm f, kin}$, we fix the reservoir threshold parameter to $f_{\rm thr} = 20$, which is the value used in the Illustris TNG simulations \citep{Weinberger2017}, and perform five runs with $\epsilon_{\rm f, kin} = 0.0025$, $0.005$, $0.008$, $0.01$, and $0.02$. To study how $f_{\rm thr}$ affects the BH growth and the star formation, we fix $\epsilon_{\rm f, kin} = 0.008$, and perform four additional runs with $f_{\rm thr} = 2$, $5$, $10$, and $40$.

In Fig.~\ref{fig:ap_kin_fb_sfr_bhar}, we present the time-evolution of the total SFRs, the total BH accretion rates, and the total BH masses for the kinetic feedback runs. In the left panels, we have fixed $f_{\rm thr} = 20$ but vary $\epsilon_{\rm f, kin}$, while in the right panels, the tests are run with fixed $\epsilon_{\rm f, kin} = 0.008$ but varying $f_{\rm thr}$. We also over-plot the results from a pure thermal feedback run with $\epsilon_{\rm f, th} = 0.02$ (black line) for comparison.

The BH accretion rates and BH masses are more sensitive to the feedback efficiency parameter. With a higher $\epsilon_{\rm f, kin}$, the total feedback energy from the radiated luminosity is higher, and the BH accretion rate is more significantly suppressed with the final BH mass being lower. By varying $\epsilon_{\rm f, kin}$ over an order of magnitude, from $0.0025$ to $0.02$, the overall BH accretion rate (and the final BH mass) also differs by roughly an order of magnitude. In contrast, the BH accretion rate and the final BH mass are less sensitive to $f_{\rm thr}$. With a higher $f_{\rm thr}$, the strength of each individual AGN feedback event is stronger, but the feedback events are less frequent. The test runs show that the latter effect tends to be more impactful for BH growth, i.e. as the feedback events are less frequent, the BHs tend to have slightly higher accretion rates. When $f_{\rm thr}$ varies from $2$ to $40$, the BH accretion rate and the final BH mass only change by a factor of ${\sim} 2$.

The star formation history is more sensitive to the reservoir threshold parameter, in agreement with the findings in \citet{Weinberger2017}. In the top left panel of Fig.~\ref{fig:ap_kin_fb_sfr_bhar}, kinetic feedback runs with different $\epsilon_{\rm f, kin}$ show quite similar star formation histories. In contrast, in the top right panel, when $f_{\rm thr}$ increases from $2$ to $40$, it is evident that the SFRs decrease by a factor of a few (especially after the first passage). This suggests that with a stronger AGN feedback pulse, it is more effective to kick out and shock-heat the gas, and thus further quenching the star formation in a galaxy.

\begin{figure}
\centering\includegraphics[width=\columnwidth]{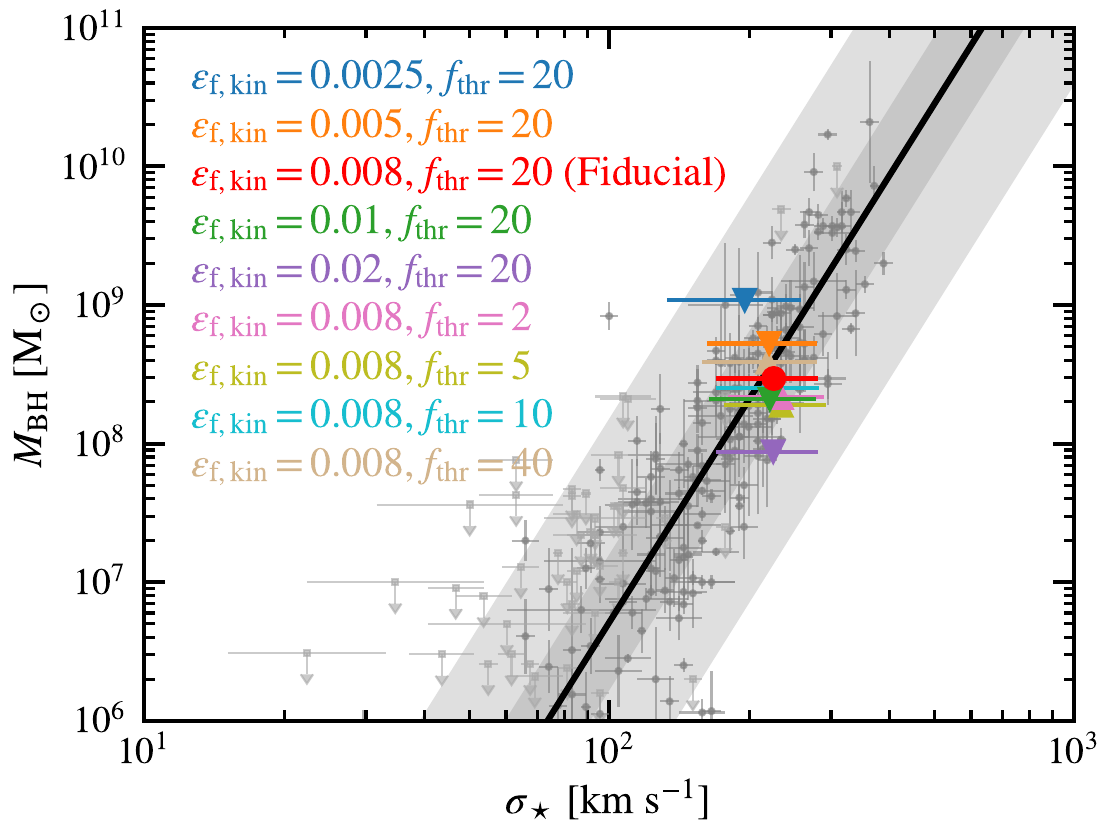}
\caption{The $M_{\rm BH}$-$\sigma_{\star}$ relation for the merger remnants from the DD-11-G0 merger runs with different kinetic feedback model parameters. The fiducial run is shown with a red filled circle. The runs with varying $\epsilon_{\rm f, kin}$ but fixed $f_{\rm thr} = 20$ are shown with downward-triangles, while the runs with varying $f_{\rm thr}$ but fixed $\epsilon_{\rm f, kin} = 0.008$ are shown with upward-triangles. The error bars show the $3\sigma$ uncertainty of the line-of-sight velocity dispersion computed from 50 random line-of-sight directions. The observational data points (grey scatter points with error bars or upper limits) and the best-fitting observed $M_{\rm BH}$-$\sigma_{\star}$ relation (black solid line) are from \citet{van_den_Bosch2016}, with the grey shaded regions denoting one and three times the intrinsic scatter.}
\label{fig:ap_kin_fb_M_sigma_relation}
\end{figure}

The aforementioned impacts of $\epsilon_{\rm f, kin}$ and $f_{\rm thr}$ on the BH growth and star formation history indicates that the following methods can be used to calibrate these parameters:

(i) We can use the scaling relations related to the BH masses, e.g. $M_{\rm BH}$--$\sigma_{\star}$ relation, to constrain $\epsilon_{\rm f, kin}$. In Fig.~\ref{fig:ap_kin_fb_M_sigma_relation}, we plot the galaxy remnants from all the kinetic feedback runs on the $M_{\rm BH}$--$\sigma_{\star}$ plane. As expected, the location of a galaxy remnant on the $M_{\rm BH}$--$\sigma_{\star}$ plane is more sensitive to $\epsilon_{\rm f, kin}$, and the runs with $\epsilon_{\rm f, kin} = 0.008$ agree well with the best-fitting observed $M_{\rm BH}$--$\sigma_{\star}$ relation from \citet{van_den_Bosch2016}. Therefore, we choose $0.008$ as the fiducial value for $\epsilon_{\rm f, kin}$ in our simulations presented in the main text.

(ii) We can use the observations of the star formation history or galaxy gas mass fractions to constrain $f_{\rm thr}$. This can be readily done in cosmological galaxy formation simulations (e.g. by comparing the evolution of the cosmic SFR density to observations), but it cannot be done with idealised galaxy merger simulations as the initial conditions are ideally generated and the galaxy samples are limited. Therefore, in this study, we simply adopt the value of $f_{\rm thr} = 20$ suggested in \citet{Weinberger2017} as our fiducial value. In the future two-mode AGN feedback model for ketju cosmological simulations, we can use the cosmic SFR density to provide a better calibration of this parameter.

\begin{figure}
\centering\includegraphics[width=\columnwidth]{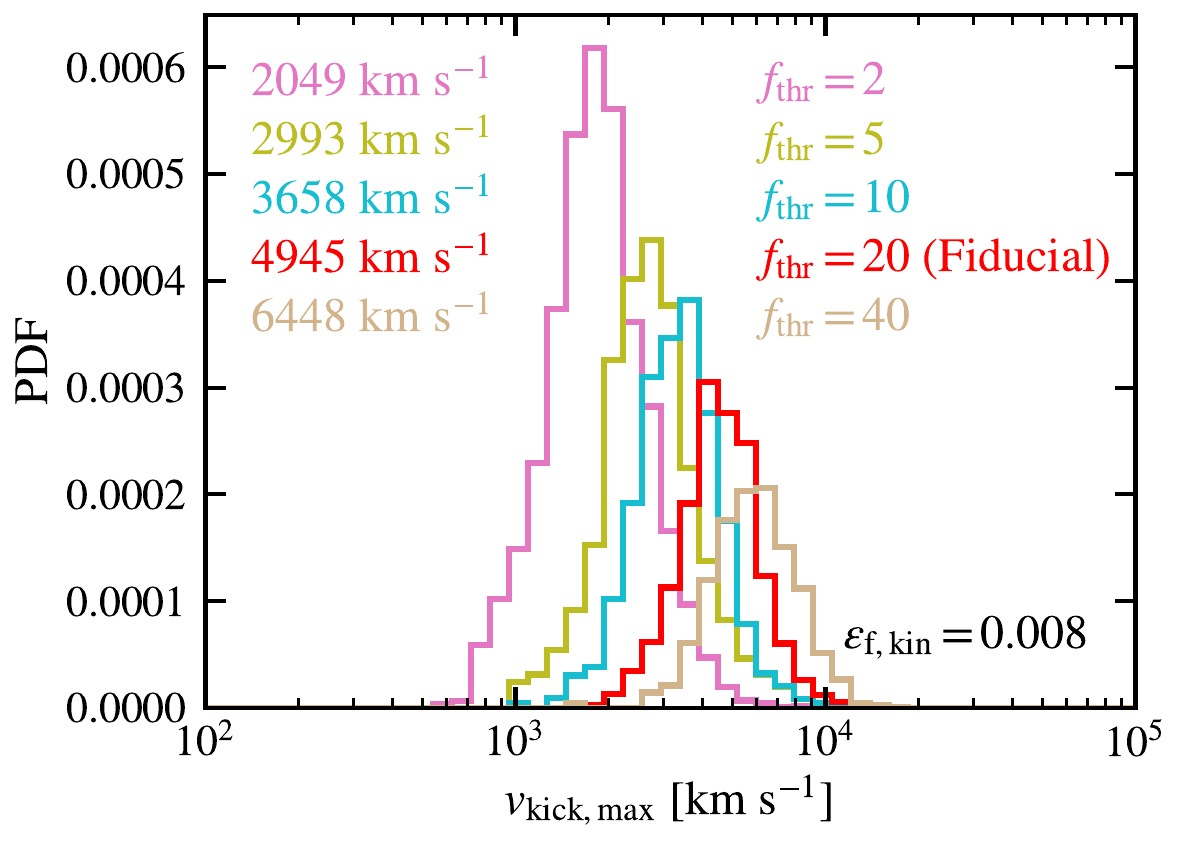}
\caption{Probability distribution function (PDF) of the maximum AGN feedback kicking velocity, $v_{\rm kick, max}$. The median $v_{\rm kick, max}$ of each distribution is given in the upper left corner.}
\label{fig:ap_kin_fb_vkick_pdf}
\end{figure}

As $f_{\rm thr}$ controls the feedback strength and thus the kick velocities of gas neighbours, it is interesting to study the gas kick velocities in runs with different $f_{\rm thr}$ (fixed $\epsilon_{\rm f, kin} = 0.008$) and compare them to the observed AGN outflow velocities. In Fig.~\ref{fig:ap_kin_fb_vkick_pdf}, for each test run, we compute the maximum kick velocity, $v_{\rm kick, max}$, from each feedback event and plot the probability distribution function (PDF) of $v_{\rm kick, max}$ from all feedback events during the whole run. As expected, with a higher $f_{\rm thr}$, the median $v_{\rm kick, max}$ is higher. With the fiducial value $f_{\rm thr} = 20$, the median $v_{\rm kick, max}$ is ${\sim} 5000~{\rm km}~{\rm s}^{-1}$ and in some feedback events, $v_{\rm kick, max}$ can reach ${\sim} 10^{4}~{\rm km}~{\rm s}^{-1}$, in agreement with the typical gas outflow velocities from observations \citep{Crenshaw2003,Laha2021}. Note that a fixed kick velocity of a few thousand to $10^{4}~{\rm km}~{\rm s}^{-1}$ has been implemented in many previous kinetic AGN feedback models \citep[see e.g.][]{Choi2012,Debuhr2012,Barai2014,Choi2014,Choi2015,Barai2016,Angles-Alcazar2017,Torrey2020}. The agreement with observations and other simulation models support the notion that our adopted fiducial value for $f_{\rm thr}$ is a reasonable choice.

In Fig.~\ref{fig:ap_kin_fb_sfr_bhar}, by comparing the pure thermal feedback and pure kinetic feedback runs, we find that they have qualitatively similar star formation and BH growth histories. Especially, our fiducial kinetic model (red lines) gives a fairly similar BH growth history as in the fiducial thermal feedback run (black lines). However, quantitatively the kinetic feedback model is more effective at quenching star formation compared to the thermal feedback model.

\section{DD-11-G5 kinetic AGN feedback runs}\label{ap:dd_11_g5}

\begin{figure*}
\centering\includegraphics[width=425pt]{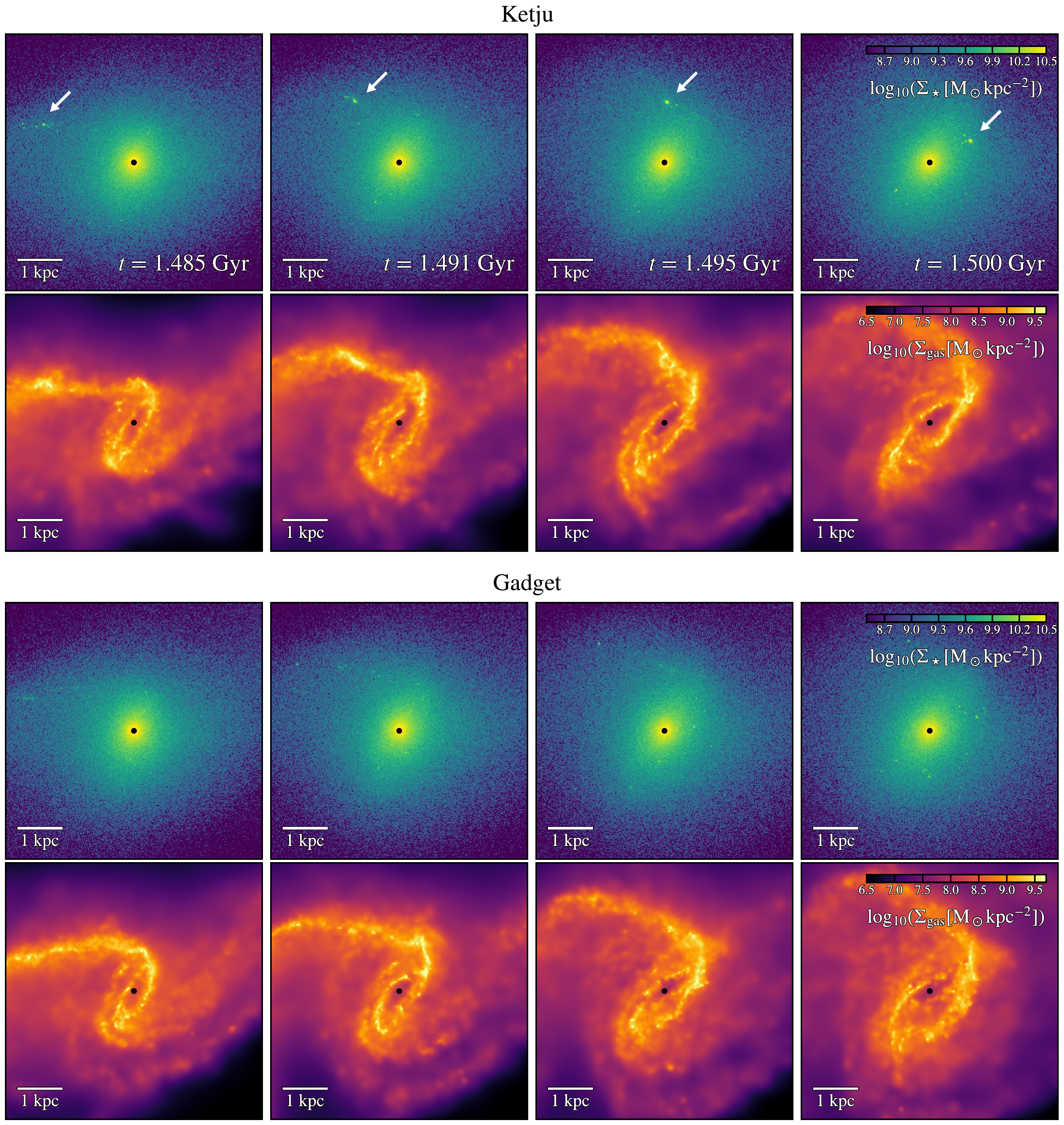}
\caption{Formation of an ultra-compact stellar clump in the ketju run. The evolution of the central stellar and gas distributions around the secondary BH (marked by black dots) in the ketju run is plotted in the first and second row respectively, and the third and fourth rows show the same for the gadget run. The distribution is projected along the $y$-axis direction and plotted on the $xz$-plane. In the first row, the targeted compact stellar clumps are indicated with white arrows. From left to right columns, the snapshots at $t = 1.485$, $1.491$, $1.495$, and $1.500$ Gyr are plotted.}
\label{fig:ap_dd_11_g5_stellar_clump}
\end{figure*}

\begin{figure*}
\centering\includegraphics[width=425pt]{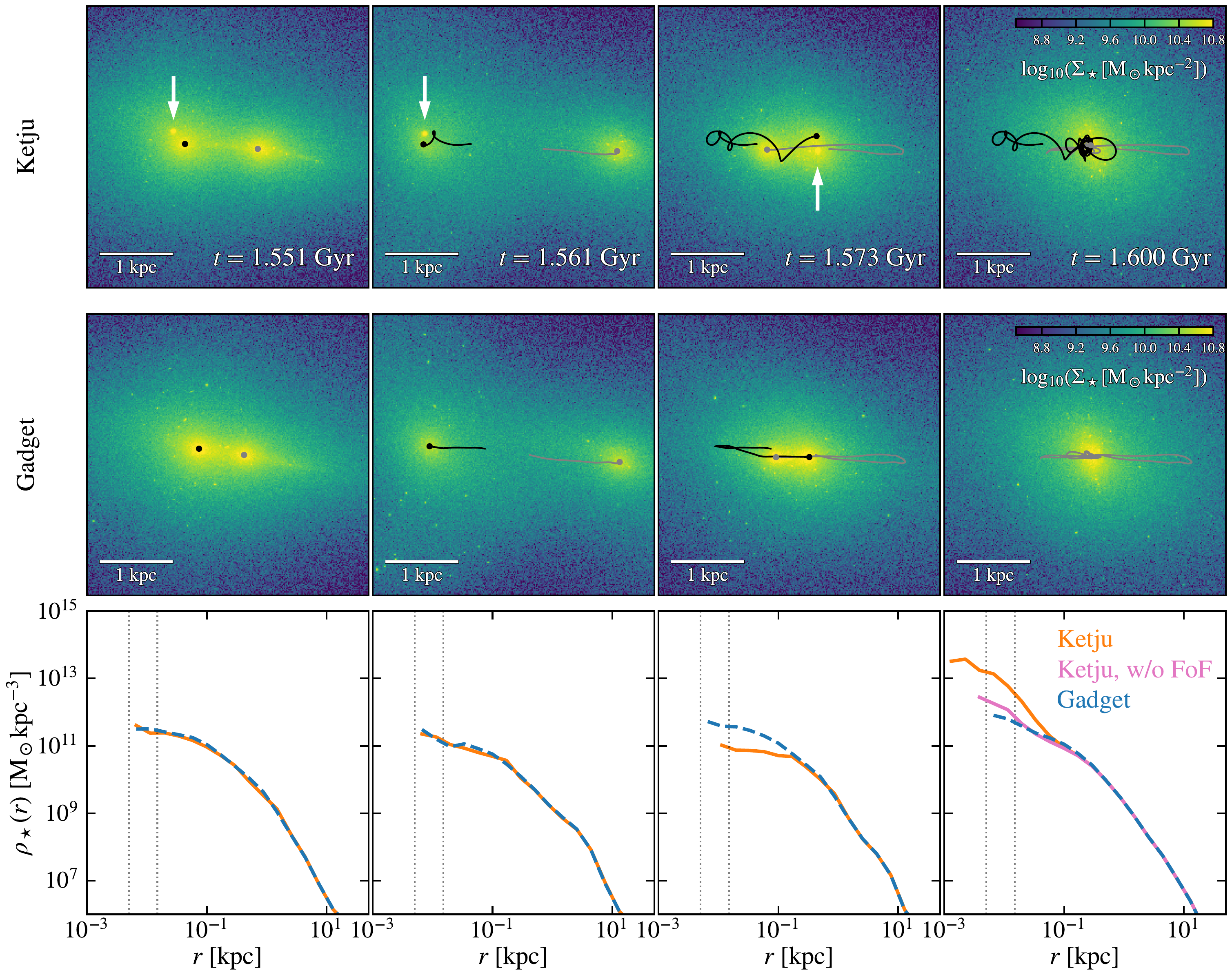}
\caption{The impact of the ultra-compact stellar clump on the BH trajectory and stellar density profile. The first row shows the evolution of the stellar distribution from $t=1.551$ to $1.600$ Gyr in the ketju run. The targeted ultra-compact stellar clump in the ketju run is indicated with white arrows in the three left panels, and in the rightmost panel this clump has already merged into the galaxy remnant centre. The secondary and primary BHs are marked by black and grey dots respectively, and their trajectories since $t=1.551$ Gyr are plotted with black and grey solid lines. The second row shows similar plots but for the gadget run. Note that in the gadget run, two BHs merge at $t = 1.584$ Gyr; after that only the primary BH (grey) is plotted (i.e. the rightmost panel). The third row shows the spherically averaged stellar density profiles centred on the secondary BH (or the merged BH in the rightmost panel for the gadget run) in both ketju (orange solid) and gadget (blue dashed) runs. The vertical dotted lines mark the stellar softening length $\epsilon_{\star}$ and the ketju radius $3\epsilon_{\star}$. In the rightmost panel, the magenta solid line shows the stellar density profile in the ketju run after removing the particles of the ultra-compact stellar clump identified using the FoF method.}
\label{fig:ap_dd_11_g5_stellar_rho}
\end{figure*}

\begin{figure}
\centering\includegraphics[width=\columnwidth]{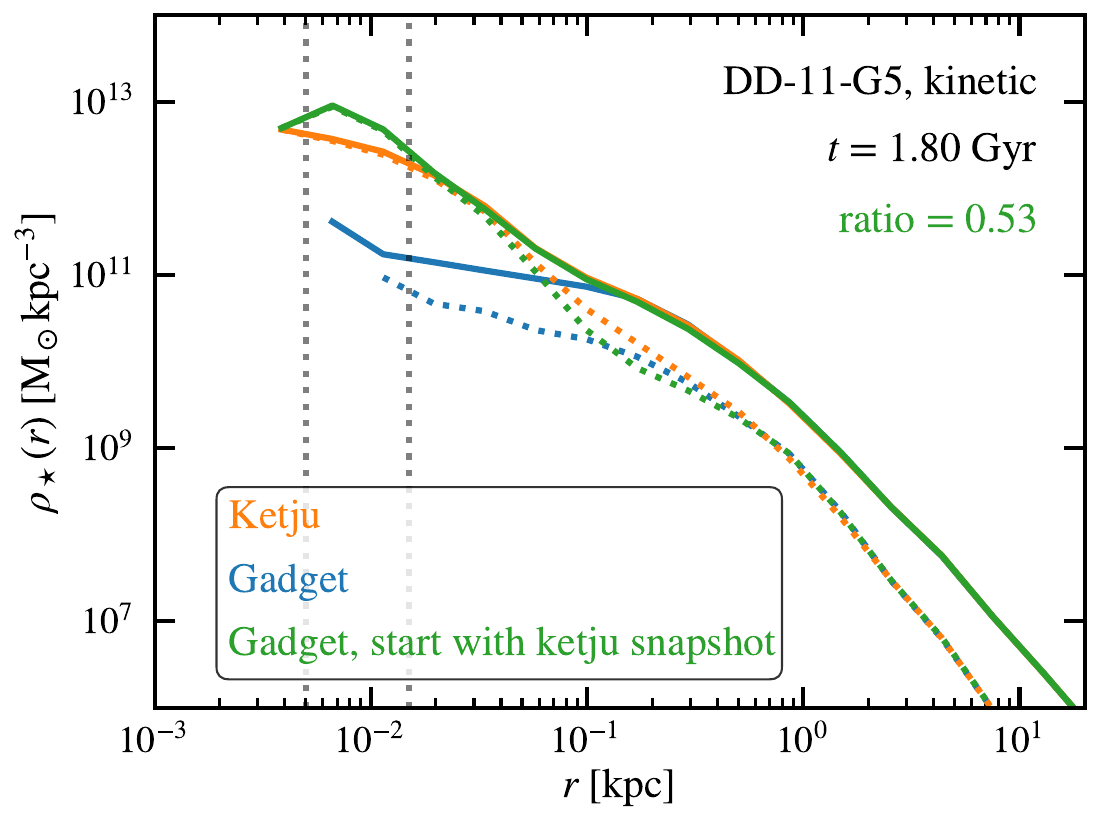}
\caption{Similar to the panel of the DD-11-G5 kinetic feedback run in Fig.~\ref{fig:density_profile_after}, but adding the profile from the gadget test run (green).}
\label{fig:ap_dd_11_g5_gadget_test}
\end{figure}

As shown in Fig.~\ref{fig:density_profile_after}, the DD-11-G5 kinetic feedback simulation is a clear outlier which exhibits significantly different behaviour compared to other simulations, i.e. at $t = 1.80$ Gyr, the ketju central stellar density is remarkably higher than the gadget density even though they are very similar at $t = 1.55$ Gyr (Fig.~\ref{fig:density_profile_before}). In this appendix, we investigate the cause of this difference. 

In order to do this, we have rerun both the ketju + binary accretion and gadget runs starting from the snapshots at $t = 1.45$ Gyr (i.e. snapshot 29 in the original runs) and increased the output frequency with an output interval of ${\sim} 2.5$ Myr (compared to the interval of ${\sim} 50$ Myr in the original runs). These two new runs continue until $t = 1.60$ Gyr, and each of them has $60$ output snapshots. These outputs with higher time resolution enable us to better examine the galaxy merging process and the BH evolution.

Fig.~\ref{fig:ap_dd_11_g5_stellar_clump} shows the evolution of the central stellar and gas distributions in both the ketju (upper two rows) and gadget (lower two rows) runs from $t=1.485$ to $1.5$ Gyr. Overall, at $t=1.485$ Gyr, the large-scale patterns of the gas tidal arms are quite similar in both runs. However, within the tidal arm, in the ketju run the gas tends to be more compact in the top-left corner while in the gadget run it tends to be smoother. This small difference in the gas tidal arms leads to minor differences in the star formation, i.e. in the ketju run the small stellar clumps are more compact and clustered (marked by the white arrow) while in the gadget run they are more diffuse and are more separately distributed along the arm. In the subsequent evolution, small stellar clumps in the ketju run merge to form an ultra-compact stellar clump orbiting the galaxy centre in which the secondary BH resides, amplifying the initial small difference in a runaway manner. In contrast, in the gadget run the stellar clumps remain diffuse and separated while moving along the tidal arm.

After $t = 1.500$ Gyr, the ultra-compact stellar clump continues to orbit the galaxy centre of the secondary BH and gradually sinks to the central region. We plot the later evolution of the stellar densities in both runs in Fig.~\ref{fig:ap_dd_11_g5_stellar_rho}.

At the snapshot with $t = 1.551$ Gyr, we identify the ultra-compact stellar clump (marked by the white arrow) using the Friends-of-Friends \citep[FoF,][]{Davis1985} method with a linking length of $5$ pc. The identified FoF clump has a total stellar mass of $4.8 \times 10^{8}~{\rm M}_{\sun}$ and a stellar half-mass radius of $16.3$ pc. Possible observational analogues of this ultra-compact clump are ultra-compact dwarfs (UCDs) for which masses are $10^{6} \la M_{\star} \la 10^{8}~{\rm M}_{\sun}$ and the effective radii are $10 \la R_{\rm e} \la 100$ pc \citep[][]{Mieske2008,Brodie2011}. In terms of mass and size, the ultra-compact clump in our simulation is similar to the most massive known UCDs located near the elliptical galaxy M59, M59-UCD3, which has a dynamical mass of a few times $10^{8}~{\rm M}_{\sun}$ and an effective radius of around $20$ pc \citep{Liu2015,Sandoval2015}. One of the possible formation mechanisms for UCDs is that they are stellar super-clusters merged from smaller star clusters formed in the tidal arms of gas-rich galaxy mergers \citep{Fellhauer2002,Fellhauer2005}, which is quite similar to the formation process seen in our simulations.

As the ultra-compact clump moves close to the galaxy centre of the secondary BH, it affects the motion of the secondary BH. The clump and the BH orbit each other, leading to the helix-like BH trajectory, as shown in the first row of Fig.~\ref{fig:ap_dd_11_g5_stellar_rho}. At $t \sim 1.59$ Gyr, this clump finally merges into the galaxy remnant and boosts the central density in the ketju run. The stellar density profiles at $t = 1.600$ Gyr for both the ketju and gadget runs are shown in the rightmost panel of the bottom row. To further confirm that the density bump in the ketju run is indeed due to the ultra-compact clump, we exclude the star particles which belong to the FoF group identified in the $t = 1.551$ Gyr snapshot and re-compute the stellar density profile which is shown with the magenta solid line. We can clearly see that after removing the contribution from the ultra-compact clump, the stellar density profile in the central region decreases significantly and agrees better with the gadget one.

As a final test, we start with the ketju snapshot at $t=1.55$ Gyr (i.e. the `before' snapshot), which contains the ultra-compact clump, and run with the gadget mode until $t = 1.80$ Gyr (the `after' snapshot). Since the effect of the ultra-compact clump is included in this gadget test run, we can better isolate the effect of star ejections when comparing it with the original ketju run. The final stellar density profile centred on the merged BH is plotted with the green curve in Fig.~\ref{fig:ap_dd_11_g5_gadget_test}. Now we can see the same behaviour as in the other simulations shown in Fig.~\ref{fig:density_profile_after} when comparing the ketju and the gadget test runs, i.e. the central stellar density in the ketju run tends to be lower than the gadget one due to the gravitational slingshot ejection. Quantitatively, the ketju-to-gadget enclosed mass ratio within the ketju radius changes from $12.49$ in the original gadget run to $0.53$ in this test gadget run. We have also performed similar tests for other galaxy merger runs in Fig.~\ref{fig:density_profile_after}, and confirmed that the central density profiles at the `after' snapshots of the test gadget runs are all higher than their ketju counterparts, further supporting our discussions in Section~\ref{subsec:density_profile}, which were based on the other galaxy merger runs. 

\label{lastpage}

\end{document}